\pdfoutput=1

\documentclass[structabstract]{aa}
\usepackage{graphicx}
\usepackage{caption}
\usepackage{subfig}
\usepackage{txfonts}
\usepackage{natbib}
\usepackage{color}
\bibpunct{(}{)}{;}{a}{}{,} % to follow the A&A style

\graphicspath{ {images/} }

\begin{document}
\title{New member candidates of Upper Scorpius from Gaia DR1}
\subtitle{}
   \author{S. Wilkinson\inst{1}
   \and B. Mer\'{i}n\inst{1}
   \and P. Riviere-Marichalar\inst{2}}
   \institute{European Space Astronomy Centre (ESA), PO Box 78, 28691 Villanueva de la Ca\~{n}ada, Spain
   \and Instituto de F\'isica Fundamental (CSIC). Calle Serrano 113b \& 123, E-28006 Madrid, Madrid, Spain
   }
   \authorrunning{S. Wilkinson}
   \titlerunning{New member candidates of Upper Scorpius from Gaia DR1}
   \date{}

\abstract {Selecting a cluster in proper motion space is an established method for identifying members of a star forming region. The first data release from Gaia (DR1) provides an extremely large and precise stellar catalogue, which when combined with the Tycho-2 catalogue gives the 2.5 million parallaxes and proper motions contained within the Tycho-Gaia Astrometric Solution (TGAS).} {We aim to identify new member candidates of the nearby Upper Scorpius subgroup of the Scorpius-Centaurus Complex within the TGAS catalogue. In doing so, we also aim to validate the use of the DBSCAN clustering algorithm on spatial and kinematic data as a robust member selection method.} {We constructed a method for member selection using a density-based clustering algorithm (DBSCAN) applied over proper motion and distance. We then applied this method to Upper Scorpius, and evaluated the results and performance of the method.} {We identified 167 member candidates of Upper Scorpius, of which 78 are new, distributed within a 10$^{\circ}$ radius from its core. These member candidates have a mean distance of 145.6 $\pm$ 7.5 pc, and a mean proper motion of (-11.4, -23.5) $\pm$ (0.7, 0.4) mas/yr. These values are consistent with measured distances and proper motions of previously identified bona-fide members of the Upper Scorpius association.} {}


\maketitle
   
\section{Introduction} 
Analysis of stellar kinematics has provided the basis for the development of multiple robust member selection methods for stellar clusters. Such methods can be used for the investigation of stellar membership across time for nearby galactic stellar clusters. This in turn allows for the measurement of cluster disk fractions as a function of cluster age, which can be used to determine the time-scale for exoplanet formation within the solar neighbourhood. These methods, therefore, can have an important impact on the development of planet formation theories. Methods such as those developed by \cite{deB1999} (an updated Convergent Point method) and \cite{HA1999} (the Spaghetti method) assume isotropic velocity dispersion about some central point, and so attempt to identify member candidates by looking for intersecting proper motion vectors. \cite{deZ1999} used both of these methods with great success on the nearby OB associations, including Upper Scorpius. \cite{Platais1998} looked for ``clumps'' in proper motion space, selecting neighbours within 2.5 mas/yr, as a starting point for a member selection method. Similarly, \cite{Lepine2009} used a proper motion selection method based on the mean proper motion of a cluster within a more sophisticated member selection method involving multiple photometric tests. Recently, \cite{Malo2013} (BANYAN) and \cite{Gagne2013} (BANYAN II) have constructed member selection techniques using Bayesian analysis on kinematic and photometric data. For many of these studies, kinematic data was obtained from the Hipparcos catalogue. The Tycho-Gaia Astrometric Solution (TGAS) \citep{michalik2015} catalogue provides significant accuracy improvements in proper motion and parallax compared with Hipparcos, providing an opportunity to revisit stellar kinematics-based member selection methods. 

\begin{figure} [ht]
    \resizebox{\hsize}{!}{\includegraphics{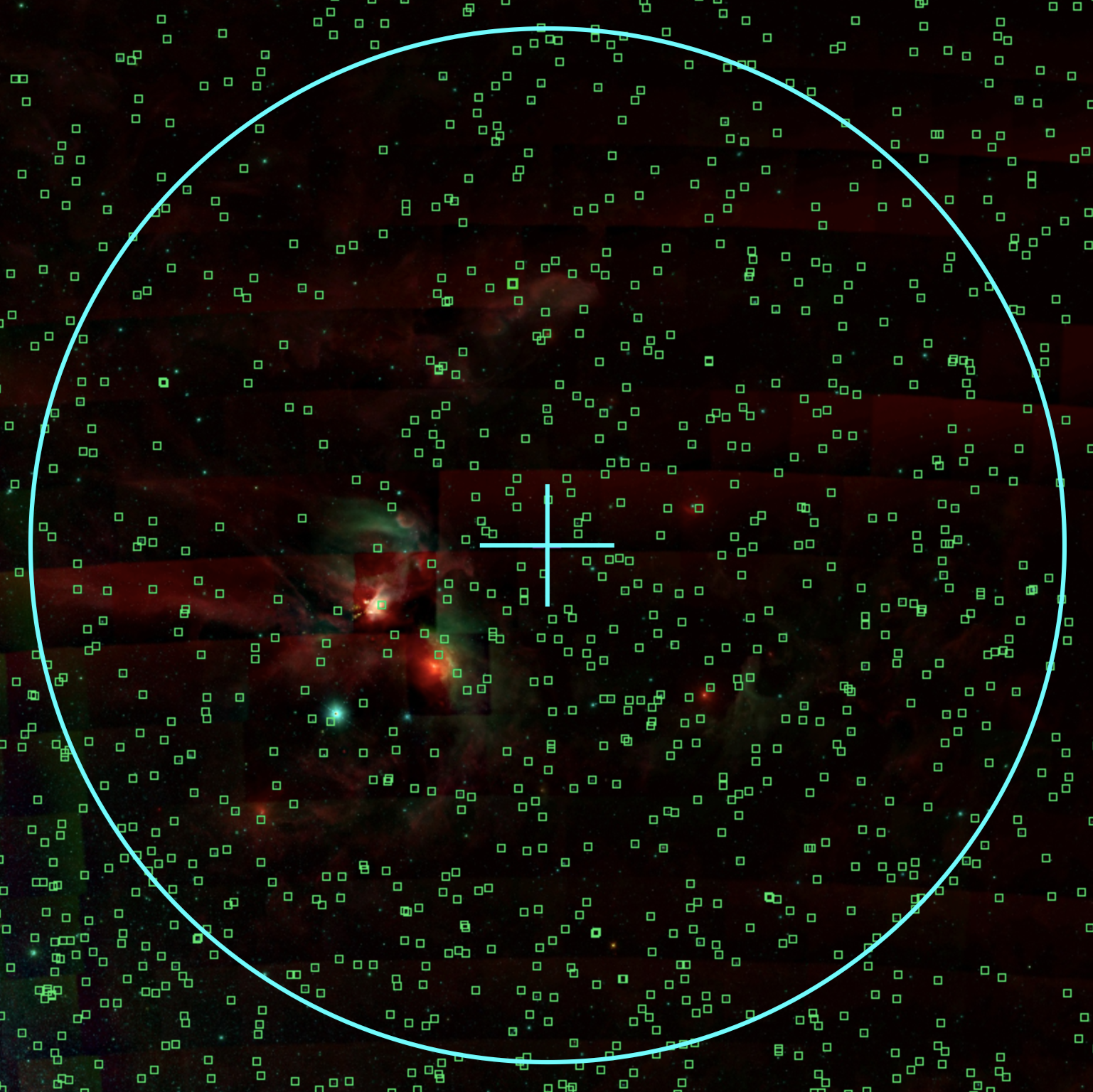}}
    \caption{WISE Color image centered on Upper Scorpius with TGAS objects overlaid, as seen in ESASky \citep{baines2016}. The circle denotes the 10$^{\circ}$ surrounding the center of Upper Scorpius. North is up, and East is left.}
    \label{fig:ESASky}
\end{figure}

In \cite{gaiacollaboration2016}, the authors used a simple proper motion clustering method in order to obtain an estimate of the mean parallax of the Pleiades cluster. They first selected all stars within a 5 degree radius of the center of the Pleiades cluster. They then required the proper motion of Pleiades members to be within 6 mas/yr of the mean proper motion determined by previous proper motion studies of the Pleiades cluster. From these member candidates they calculated the mean parallax. While this method provides a reasonable estimate of the mean parallax of the Pleiades, it has its limitations. This method assumes perfect spherical symmetry in proper motion space, so there is a strong likelihood that the member candidates include many false positives. The core concept of selecting a cluster in proper motion space with constrained position has merit, one expects members of an association to be co-moving and close together. We believe the potential of clustering algorithms in member selection is strong, especially given the accuracy improvements provided by TGAS. However, a more sophisticated algorithm is required for a more robust member selection method.

Here we present such a method, using the DBSCAN algorithm \citep{ester1996} to select a cluster in proper motion and distance. The advantage of DBSCAN for this application over other clustering algorithms such as k-means via Principal Component Analysis (PCA), is that it is a purely density-based clustering algorithm. Accordingly, it can identify clusters of arbitrary shape. Since we are simply looking for an over-density in a large field of background objects, DBSCAN is well suited. We applied our method to the Upper Scorpius subgroup of the Scorpius-Centaurus Complex. From this we identified several member candidates of Upper Scorpius, and obtained an estimate for the mean distance and proper motion of those member candidates. These estimates were then compared with estimates from previous studies of Upper Scorpius. The spatial distribution of the member candidates was also qualitatively analysed.

\section{Data Analysis}

We began our analysis with the 11232 objects, shown in Figure \ref{fig:ESASky}, present in the TGAS catalogue \citep{michalik2015} within a cone of radius 10$^{\circ}$ centered at a right ascension of 16h 11m 60.00s (J2000) and a declination of -23$^{\circ}$ 23' 60.00'' (J2000). This region and the TGAS objects within it are shown in Figure \ref{fig:ESASky}. As described in Section 3, a 10$^{\circ}$ radius is consistent with the age and dispersion in proper motion of Upper Scorpius. These objects were accessed using the Gaia Archive, hosted at ESAC (http://archives.esac.esa.int/gaia). Those objects with relative parallax error greater than 0.1 ($\sigma_{\pi} / \pi > 0.1$) were removed from the data set. Such a filter was used by \cite{Binney1998} to ensure the accuracy of astrometric measurements. The filtered data set, upon which the clustering analysis was performed, had 2827 objects. Given that nearly 8400 objects were removed by the filter, without it a significant portion (75\%) of the data used in the analysis would have had poor parallax accuracy. The distance histograms of the sample with and without the relative parallax error filter were compared. The filter introduced a bias towards more nearby objects, shifting the peak of the distance distribution from $\sim$250pc to $\sim$180pc. While the distribution for the unfiltered sample had a long tail up into the $\sim$1000pc range, the maximum distance in the filtered data was $\sim$450pc. 

The clustering analysis was performed using the DBSCAN algorithm as implemented in scikit-learn \citep{pedregosa2011}. The source code used in our analysis can be found in \cite{Wilkinson2017}. At a high level, DBSCAN first identifies so-called ``core points'' at the center of probable clusters, then expands outward from these points until it reaches low-density noise. The algorithm uses two main parameters $\varepsilon$ (eps) and minPts. Core points are defined as those with at least minPts neighbours within a radius of $\varepsilon$. Additional ``density-reachable'' points are added to a cluster if they are within a radius of $\varepsilon$ (eps) from a core point of that cluster.

The dimensions used in DBSCAN were proper motion in right ascension and declination, and distance, since we are looking for clusters of co-moving stars at similar distances. The data in each dimension of the data set was standardised such that it had a mean of 0 and a variance of 1 for the clustering analysis. This allows the data in different dimensions to be compared. DBSCAN requires a distance metric be set, in addition to $\varepsilon$ (eps) and minPts. A Euclidian metric was used, which is the scikit-learn default.

\cite{ester1996} describe a simple method to determine suitable values of \textit{minPts} and  \textit{eps}, using a sorted k-distance graph. A sorted k-distance graph shows the distance to the k\textsuperscript{th} nearest neighbour for each point, in descending order. In this method, \textit{minPts} is decided (from understanding of the data or otherwise), and the k-distance graph is generated with k set as \textit{minPts}. This graph should have a ``valley'' at the threshold point between the noise and the clusters. The value of the sorted k-distance graph at the first point at the bottom of this valley is then taken as the \textit{eps}.

\begin{figure} [ht]
    \resizebox{\hsize}{!}{\includegraphics{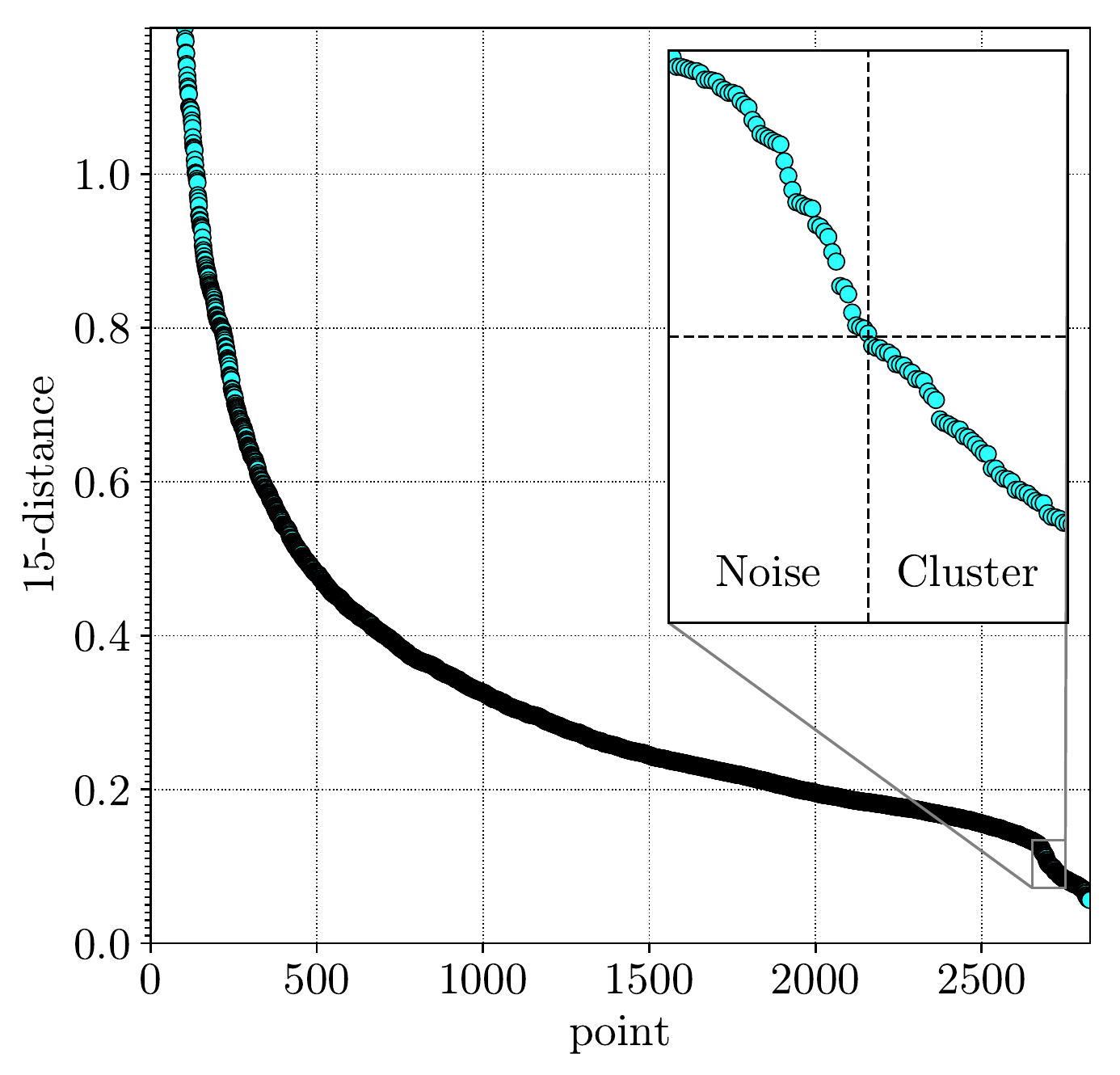}}
    \caption{Sorted 15-distance graph, with a zoomed view of the threshold point between noise (left) and the cluster (right) inset.}
    \label{fig:kdist}
\end{figure}

We set \textit{minPts} to 15, in order to minimise false positive clusters. In Appendix B we explore the behaviour of DBSCAN for other values of \textit{minPts}. A suitable value of the \textit{eps} for this value of \textit{minPts} was then determined using the method described above. Figure \ref{fig:kdist} shows the sorted 15-distance graph obtained for the 2839 objects in the filtered data set, with a zoomed view of the valley inset. We compute the 15-distance in the dimensions we are clustering over: proper motion in right ascension and declination, and distance. A valley is clearly visible at around the 2700\textsuperscript{th} point. We took the threshold point to be at the bottom of this valley. This gave us an \textit{eps} of 1.103. Note that this threshold point is very far to the right of this graph since there is a large quantity of noise (i.e. objects not in the Upper Scorpius cluster) present.

We validated this observed threshold point by using a similar method to \cite{gaiacollaboration2016} to manually select a cluster in proper motion space, thereby estimating the size of the cluster associated with Upper Scorpius. This allowed us to estimate what percentage of the data is noise, and therefore where the valley in the 15-distance graph associated with the threshold point should be observed. 

\begin{figure} [ht]
    \resizebox{\hsize}{!}{\includegraphics{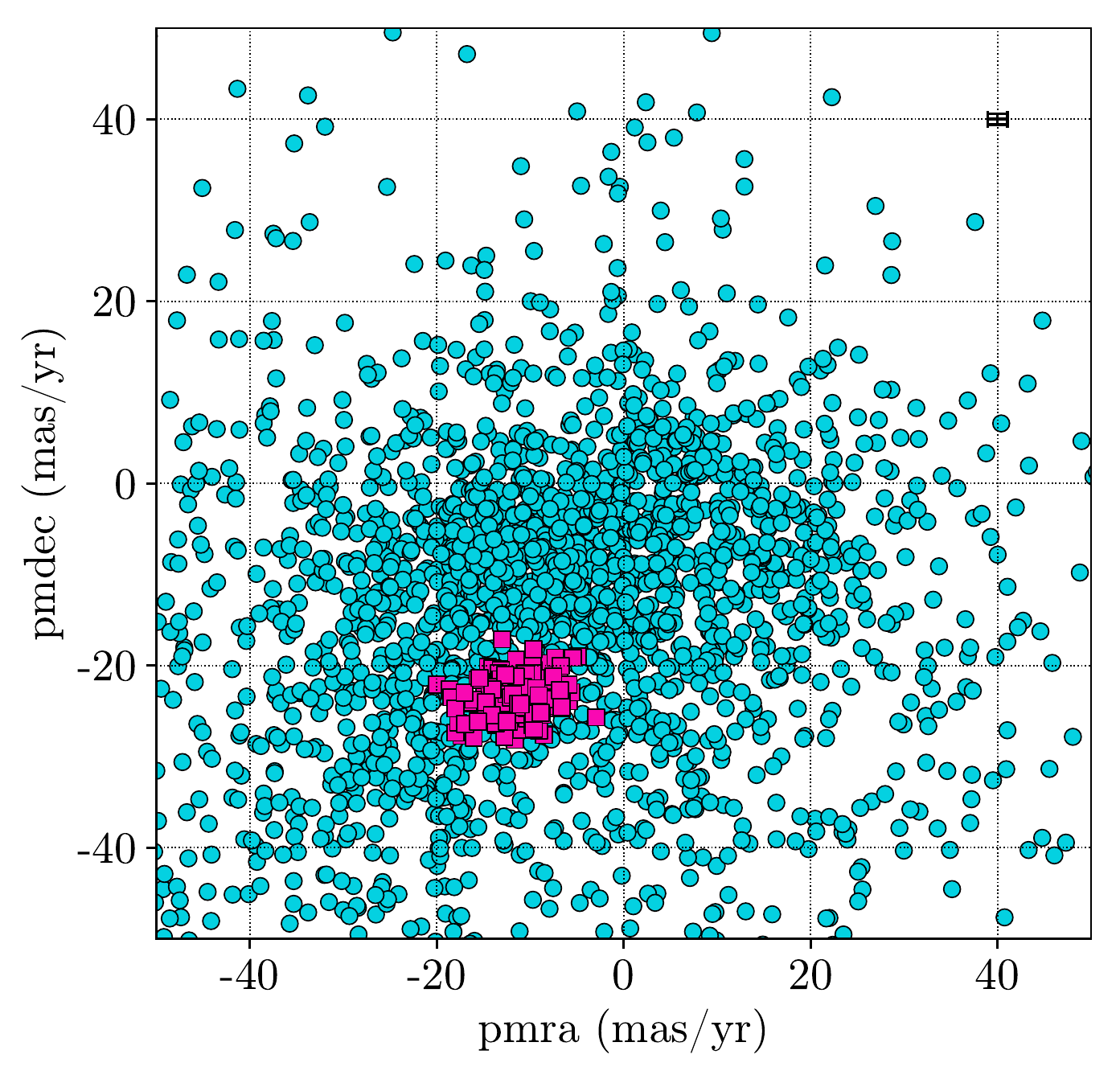}}
    \caption{Proper motion plot, with members of the cluster selected by DBSCAN shown as pink squares.}
    \label{fig:pmpm}
\end{figure}

We plotted the data in proper motion space using TOPCAT \citep{TOPCAT}, and observed a rough cluster at a proper motion of around -10 mas/yr in right ascension and a -24 mas/yr in declination. The cluster was selected by manually drawing a boundary in TOPCAT. The selected cluster was observed to have 270 members, which gives us an estimate for the percentage of noise (i.e. objects not in Upper Scorpius) of 90\%. We therefore expect to see the threshold point at around the 2600\textsuperscript{th} point, which is as observed. Note that the observed threshold point will be further to the right, since we expect to select fewer points when clustering with distance in addition to proper motion. Given that clustering is being performed over distance as well as proper motion, we are likely to be removing false positives compared to proper motion analysis alone.

\begin{figure} [ht]
    \resizebox{\hsize}{!}{\includegraphics{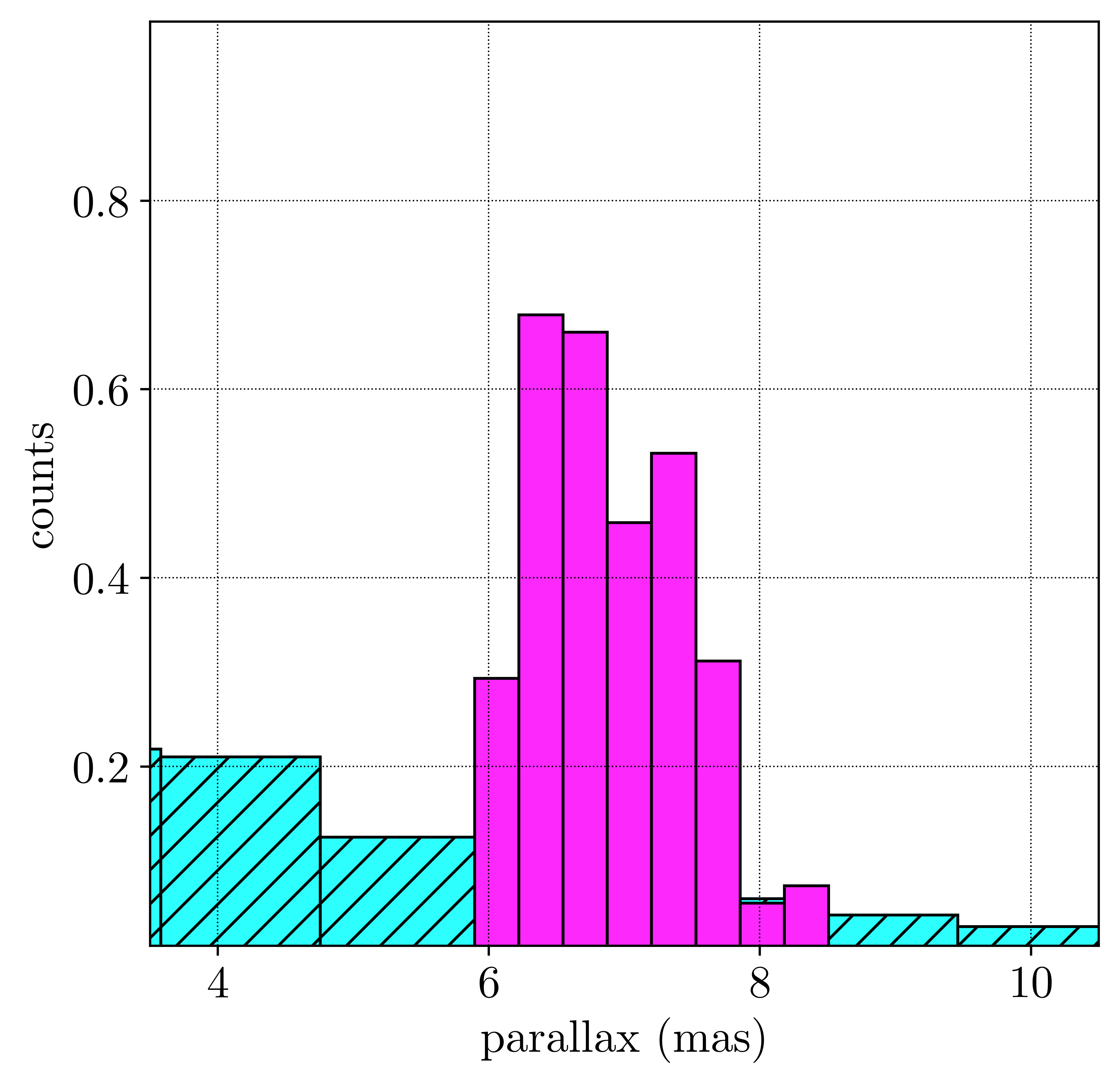}}
    \caption{Parallax histogram for all objects (blue, hatched), and the members of the cluster selected by DBSCAN (pink)}
    \label{fig:hist}
\end{figure}

Using an \textit{eps} of 0.103 and a \textit{minPts} of 15, we ran DBSCAN over the 2827 objects in the filtered data set. From this data set, DBSCAN selected 1 cluster with 167 members. These 167 member candidates are given in Table \ref{table:members}. Figure \ref{fig:pmpm} and \ref{fig:radec} show this cluster in proper motion space and in the sky plane. Figure \ref{fig:hist} shows the distribution of the parallaxes for the filtered sample of 2827 objects, and for the cluster. We used Scott's rule \citep{robitaille2013} to calculate suitable bin widths for the histogram. The cluster is very well defined in proper motion space, and in terms of parallax. The median parallax of the cluster is 6.84 mas, with a mean relative parallax error ($\sigma_{\pi} / \pi$) of 0.0947.

The mean distance of the member candidates 145.9 $\pm$ 7.5 pc. Note that here 7.5 pc is the standard error of the mean. The distance distribution of the member candidates extends from $\sim$120-165 pc, suggesting a spread of $\sim$45 pc in distance. We also obtain a mean proper motion of (-11.4, -23.5) $\pm$ (0.7, 0.4) mas/yr. \cite{deZ1999} estimates the distance to Upper Scorpius at 145 $\pm$ 2.5 pc and the mean proper motion to be (-8.1, -24.5) $\pm$ (0.1, 0.1) mas/yr. \cite{Preibisch2008} suggest a maximum spread in distance of $\sim$50 pc. \cite{Galli2018}, using data from Gaia DR1, obtain a mean distance of 146 $\pm$ 3 $\pm$ 6 pc. The consistency of the member candidates with these estimates suggests a high probability of membership in Upper Scorpius.

In order to identify the effect of brightness binning on the mean distance of the cluster, the clustering was re-run on sub-samples of the data with G-band mean magnitude above and below 10 mag. Since the vast majority of the candidate members had a magnitude greater than 10 mag (128 out of 167), no cluster was selected for the sub-sample with a magnitude less than 10 mag. For the sub-sample with magnitude greater than 10 mag, a cluster of 127 objects was selected with a mean distance of 146.1 pc. This is nearly identical to the mean distance for candidate members. This shows that any potential bias in the mean distance of the member candidates, and therefore in member selection, is not severe.

\section{Discussion}
\begin{table*}[ht]
\label{table:NUMBER}
\centering\begin{tabular}{|c|c|c|c|c|c|c|c|c|c|c|c|c|}
\hline
 Object Name & parallax & ra & dec & pmra & pmdec & US & UCL\\
 \hline
 - & mas & [h m s] $\pm$ mas & [$^{\circ}$ ' ''] $\pm$ mas & mas/yr & mas/yr & - & -\\
  \hline\hline
 HD 144569               & 7.08 $\pm$ 0.47           & 16 07 04.66 $\pm$ 0.44     & -16 56 36.06 $\pm$ 0.21      & -10.52 $\pm$ 0.05       & -20.14 $\pm$ 0.03        & 1 &    \\
 HD 144586               & 6.85 $\pm$ 0.30           & 16 07 14.92 $\pm$ 0.26     & -17 56 10.06 $\pm$ 0.15      & -8.17  $\pm$ 0.08       & -21.66 $\pm$ 0.05        & 1 &  \\
 HD 147104               & 8.09 $\pm$ 0.29           & 16 20 30.55 $\pm$ 0.27     & -20 06 52.40 $\pm$ 0.17      & -12.29 $\pm$ 0.18       & -25.16 $\pm$ 0.12        &   &   \\
 CCDM J16205-2007AB      & 8.51 $\pm$ 0.49           & 16 20 30.25 $\pm$ 0.42     & -20 07 04.24 $\pm$ 0.25      & -11.07 $\pm$ 0.20       & -26.53 $\pm$ 0.12        &   &   \\
 V933 Sco             & 7.81 $\pm$ 0.33           & 16 20 05.48 $\pm$ 0.32     & -20 03 23.40 $\pm$ 0.17      & -11.63 $\pm$ 0.05       & -24.71 $\pm$ 0.03        & 1  &   \\
 2MASS J16181997-2005348 & 7.54 $\pm$ 0.52           & 16 18 19.96 $\pm$ 0.79     & -20 05 35.20 $\pm$ 0.32      & -11.82 $\pm$ 1.79       & -24.14 $\pm$ 0.80        & 4   &    \\
 HD 145998               & 6.64 $\pm$ 0.22           & 16 14 40.14 $\pm$ 0.28     & -20 14 03.42 $\pm$ 0.17      & -17.35 $\pm$ 0.16       & -27.72 $\pm$ 0.12        & 1;3 &    \\
 HD 146366               & 7.63 $\pm$ 0.48           & 16 16 54.56 $\pm$ 0.63     & -21 37 15.86 $\pm$ 0.37      & -10.56 $\pm$ 1.34       & -24.00 $\pm$ 0.91        &   &   \\
 HD 147083               & 8.18 $\pm$ 0.80           & 16 20 28.12 $\pm$ 0.72     & -21 30 32.87 $\pm$ 0.14      & -12.70 $\pm$ 0.06       & -25.80 $\pm$ 0.04        & 1   &  \\
 HD 146416               & 7.76 $\pm$ 0.60           & 16 16 58.75 $\pm$ 0.44     & -21 18 15.26 $\pm$ 0.28      & -15.22 $\pm$ 0.04       & -25.69 $\pm$ 0.02        & 1 &    \\
 ...                     & ...      & ...            & ...           & ...      & ...            & ...   & ... \\
\hline
\end{tabular}

\caption{Member candidates of Upper Scorpius, including their SIMBAD ID. The first 10 are shown here, with the remaining available in electronic form. The ``US'' column corresponds to previously identified members of Upper Scorpius, and the ``UCL'' column corresponds to previously identified members of Upper Centaurus-Lupus. 1: \cite{Pecaut2012}, 2: \cite{Luhman2012}, 3: \cite{Chen2012}, 4: \cite{Rizzuto2015}, 5: \cite{PM2016}.}
\label{table:members}
\end{table*}

\begin{figure} [ht]
    \resizebox{\hsize}{!}{\includegraphics{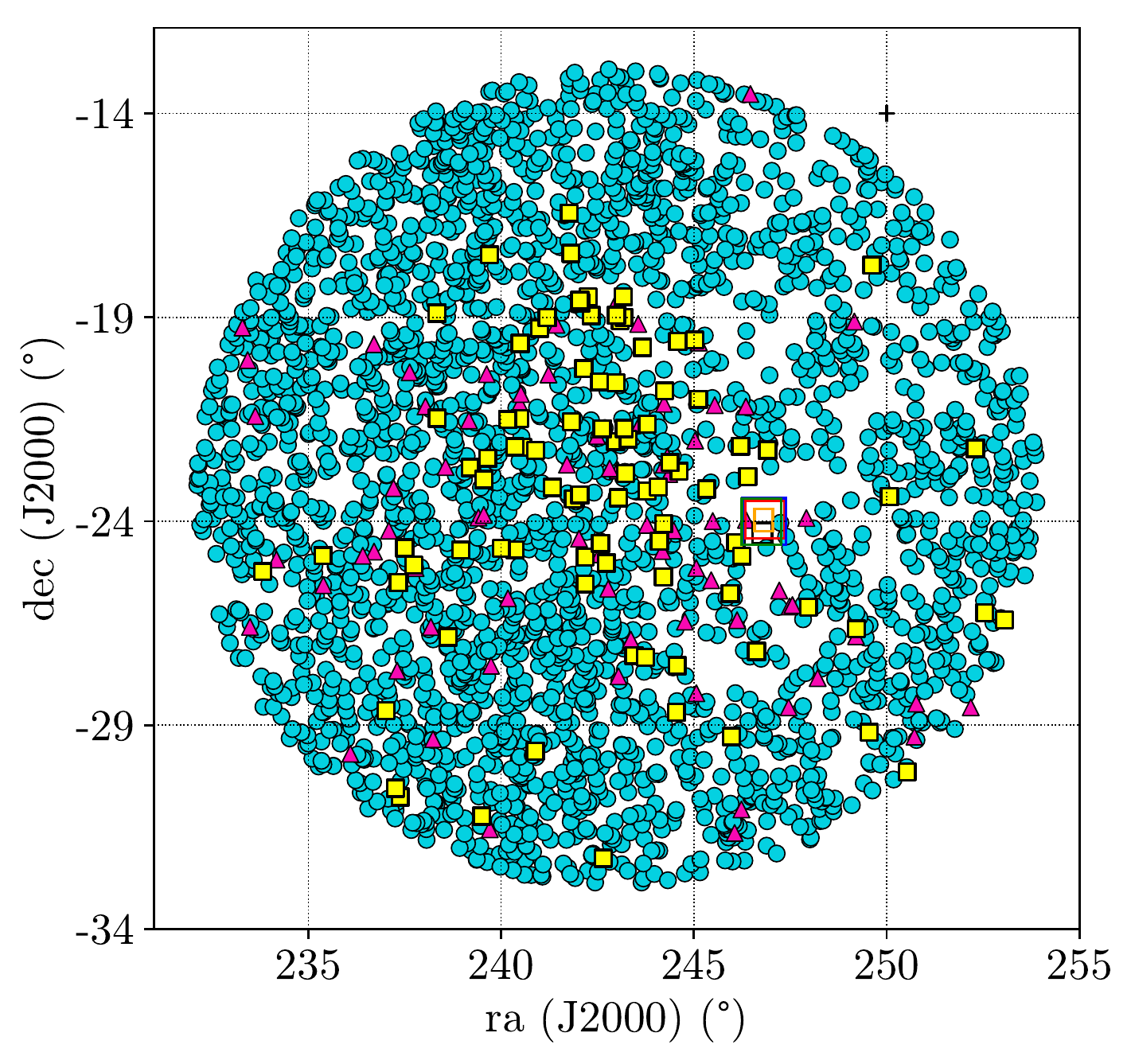}}
    \caption{Plot of the sky plane, with newly identified member candidates shown as pink triangles, and previously identified members of Scorpius-Centaurus shown as yellow squares. The boxes in the center of the plot show the extent of several previous membership studies of $\rho$-Ophiuchi. Orange: \cite{natta2002}, Green: \cite{wilking2005}, Red: \cite{alvesdeoliveira2010}, Blue: \cite{erickson2011}, Black: \cite{ducourant2017}. North is up, and East is left.}
    \label{fig:radec}
\end{figure}

We first cross-matched the member candidates with the SIMBAD catalogue, using a radius of 2 arcsec. This found corresponding SIMBAD objects for each member candidate, and accordingly the object name is included as a field in Figure \ref{table:members}. We then cross-matched the member candidates with members of Scorpius-Centaurus from \cite{Chen2012}, \cite{Pecaut2012}, \cite{Luhman2012}, \cite{Rizzuto2015}, and \cite{PM2016}. Of the 167 member candidates, 89 of them are previously identified members of Upper Scorpius. Additionally, 2 are previously identified members of Upper Centaurus-Lupus. This gives us 78 newly identified member candidates of Upper Scorpius. Given that the mean distance and mean proper motion of the candidate members are consistent with existing estimates, we are confident that we have selected strong member candidates of Upper Scorpius.

\begin{figure} [ht]
    \resizebox{\hsize}{!}{\includegraphics{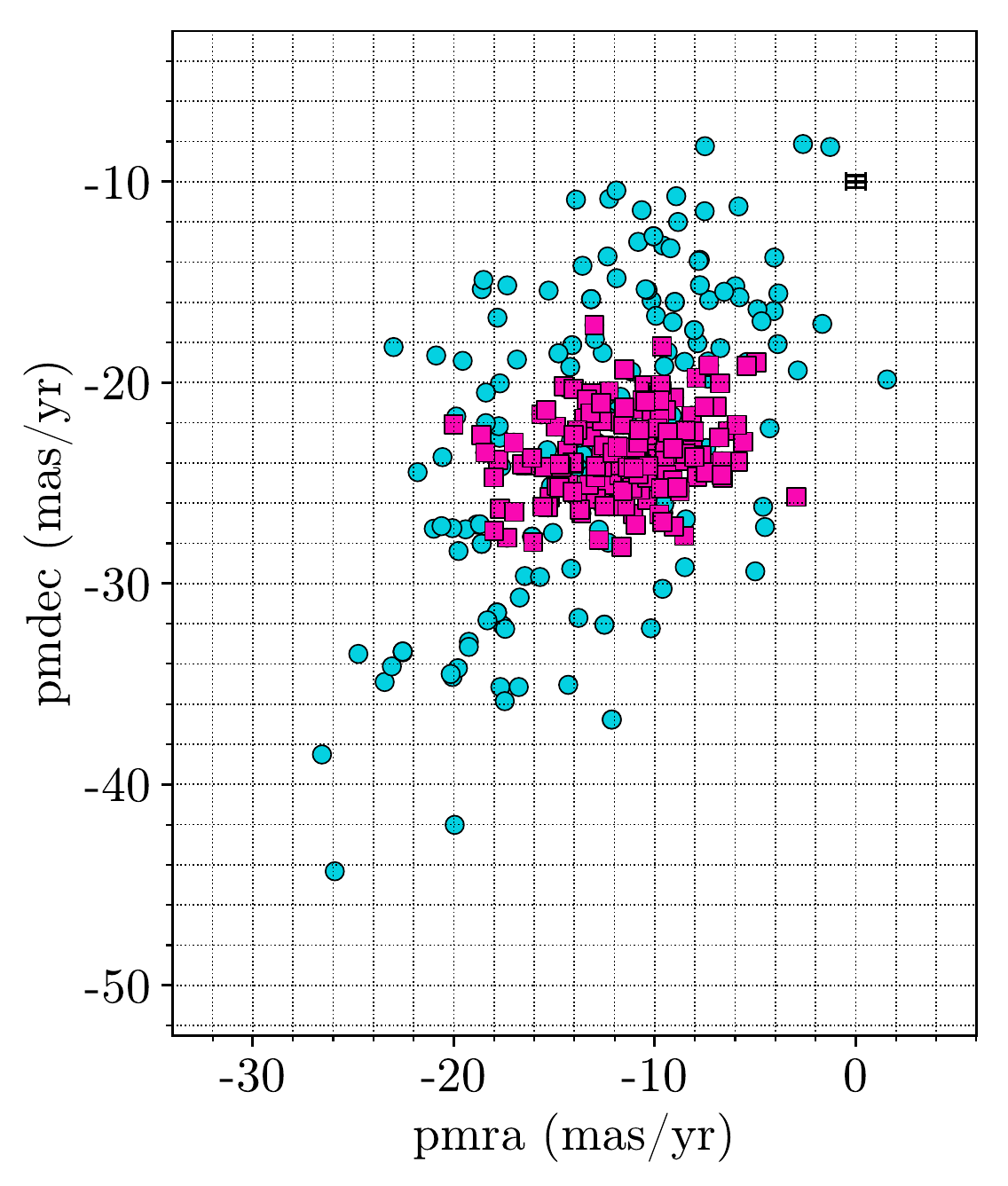}}
    \caption{Proper motion plot, with members of the cluster selected by DBSCAN shown as pink squares. Previously identified members of Upper Scorpius from \cite{deZ1999}, \cite{HA1999}, and \cite{Rizzuto2011} that are also present in the TGAS catalogue, are show as blue circles.}
    \label{fig:pmpm_overlay}
\end{figure}

We then compared the member candidates with previous stellar kinematics-based member selection methods applied to Upper Scorpius. Cross-matching the 167 member candidates with  those from \cite{deZ1999} found 55 members in common. Of the members identified by \cite{deZ1999}, 80 are present in TGAS. Similarly, cross-matching with \cite{HA1999} found 107 members in common. Of the members identified by \cite{HA1999}, 244 are present in TGAS. Cross-matching with \cite{Rizzuto2011} found 49 members in common. Of the members identified by \cite{Rizzuto2011}, 74 are present in TGAS. Figure \ref{fig:pmpm_overlay} compares the proper motion distribution of the member candidates with members previously identified by \cite{deZ1999}, \cite{HA1999}, and \cite{Rizzuto2011} that are present within TGAS and in the target region. This shows that DBSCAN can successfully recover the core of the proper motion distribution of a cluster, but struggles at the low-density extremes. This is understandable, since outside the core of the distribution the background stars provide a substantial amount of noise. Figure \ref{fig:pmpm_overlay} suggests that those objects not re-selected by DBSCAN are outside this core, which explains the discrepancy in re-selection of previous kinematically selected members. As a result, detailed analysis of the kinematic structure of the selected member candidates is not appropriate since we are selecting members from the core of the velocity distribution. The newly selected member candidates were possibly missed by previous kinematic analyses due to the systematically worse parallax uncertainties of Hipparcos (median of 0.97 mas) compared to TGAS (median of 0.32 mas). The unique nature of the DBSCAN algorithm as compared with previous methods, as described in Section 1, may have also played a role. It is worth noting that while the proper motion uncertainties of Hipparcos (median of (0.88, 0.74) mas/yr) are better than TGAS (median of (1.32, 1.32) mas/yr), the uncertainties of both the Hipparcos subset of TGAS (median of (0.07, 0.07) mas/yr) and the member candidates (median of (0.51, 0.39) mas/yr) are systematically better than both.

From Figure \ref{fig:pmpm}, we can see that the cluster of candidate members has a width of approximately 5 mas/yr in proper motion space. Two objects with a relative proper motion of 5 mas/yr would end up 10$^{\circ}$ apart after 7.2 Myr. Therefore the radius of the cone used for the selection of data from the TGAS catalogue will likely include the entirety of Upper Scorpius, given the probable young age of the region ($\sim$10 Myr, \cite{Pecaut2012}; \cite{Feiden2016}).

Upper Scorpius is in very close proximity to the $\rho$-Ophiuchus molecular cloud \citep{Preibisch2008}, so there is the potential for $\rho$-Ophiuchus members contaminating the sample. However, Figure \ref{fig:radec} shows that none of the member candidates of Upper Scorpius selected by DBSCAN are present in the areas of $\rho$-Ophiuchi observed by \cite{natta2002}, \cite{wilking2005}, \cite{alvesdeoliveira2010}, or \cite{erickson2011}, so there will be no common objects. 

The distribution of the distances within the member candidates is possibly bimodal with apparent peaks at approximately 154 pc (6.5 mas) and 133 pc (7.5 mas). While we could not establish the significance of this, a more complete sample of members may show significant bimodality. It should also be noted that there are very few member candidates that lie in the upper-right (north-west) of the sky plane of the region studied. Previous work has suggested possible substructure within the subgroups of Scorpius-Centaurus. \cite{Rizzuto2011} could not determine non-arbitrary boundaries between the subgroups of Scorpius-Centaurus due to blurring from velocity dispersion. Through producing an age map of Scorpius-Centaurus, \cite{PM2016} found potential evidence of substructure within the older subgroups of the complex (Upper Centaurus-Lupus and Lower Centaurus-Crux). Using kinematic data from Gaia DR1 and Hipparcos, \cite{Wright2018} found evidence of kinematic substructure within the subgroups of Scorpius-Centaurus. Additionally, the 3D structure of Upper Scorpius obtained by \cite{Galli2018} from Gaia DR1 supports bimodality. Further investigation of this substructure could prove fruitful, especially with data from Gaia DR2.

In Appendix A, we discuss the performance of DBSCAN applied to two sets of 20 simulations, one set produced by binning the data and uniformly generating data within those bins, and the other produced by modelling the cluster and the background as Gaussian. We estimated the number of likely contaminants to be $\sim$6-17 ($\sim$3.8-10.2\%). 

\section{Conclusions}
We applied DBSCAN to the set of TGAS objects within a 10$^{\circ}$ radius surrounding Upper Scorpius, selecting 167 candidate members. Of these member candidates, 89 are previously identified members of Upper Scorpius and 2 are previously identified members of Upper Centaurus-Lupus. The member candidates have a mean distance of 145.6 $\pm$ 7.5 pc, a spread of $\sim$45 pc in distance, and a proper motion of (-11.4, -23.5) $\pm$ (0.7, 0.4) mas/yr consistent with previous estimates for Upper-Scorpius. This suggests a strong likelihood of true membership of Upper Scorpius. In Appendix A we estimated the number of likely contaminants to be $\sim$6-17 ($\sim$3.8-10.2\%). 

While the distance distribution of member candidates suggests possible kinematic substructure in Upper Scorpius, the significance of this finding has not yet been established. Unlike many established stellar kinematics-based member selection methods, DBSCAN does not attempt to model the internal motion of associations, and relies on density alone. With the increase in precision of parallaxes and proper motions from Gaia, non-trivial substructure within associations can be investigated effectively. As a density-based clustering algorithm, DBSCAN is resilient to these substructures, so is well suited to identify them. However, while it can effectively recover the core of the proper motion distribution of an association, it struggles at the low-density fringes compared to established methods.

Analysing the color, geometric, and spectral energy distributions of the new candidate members could prove to be fruitful, but these paths are beyond the scope of this work. The 78 new candidate members should be followed up spectroscopically for their true membership to be confirmed. Were some of the new candidate members to be confirmed as true members, they would be prime candidates for ALMA imaging searches for proto-exoplanets embedded in their potential protoplanetary disks.

\begin{acknowledgements}
This work has made use of data from the European Space Agency (ESA) mission Gaia (http://www.cosmos.esa.int/gaia), processed by the Gaia Data Processing and Analysis Consortium (DPAC, http://www.cosmos.esa.int/web/gaia/dpac/consortium). Funding for the DPAC has been provided by national institutions, in particular the institutions participating in the Gaia Multilateral Agreement. This work has made use of ESASky, developed by the ESAC Science Data Centre (ESDC) team and maintained alongside other ESA science mission's archives at ESA's European Space Astronomy Centre (ESAC, Madrid, Spain). This research made use of matplotlib, a Python library for publication quality graphics \citep{hunter2007}. This research made use of NumPy \citep{van2011}. This research made use of SciPy \citep{jones2001}. This research made use of Scikit-learn \citep{pedregosa2011}. This research made use of Astropy, a community-developed core Python package for Astronomy \citep{astropy2013}. This research made use of TOPCAT, an interactive graphical viewer and editor for tabular data \citep{TOPCAT}. This work has made use of the SIMBAD database, operated at CDS, Strasbourg, France. This research has made use of the VizieR catalogue access tool, CDS, Strasbourg, France \citep{VIZIER}.
\end{acknowledgements}

\bibliographystyle{aa} % style aa.bst 2
\bibliography{biblio.bib}

\begin{appendix}

\section{Cluster Simulation and Recovery}

\addtolength{\tabcolsep}{-4.5pt} 

\begin{table*}[h!]
\centering
  \begin{tabular}{ccc}
  \includegraphics[width=57.75mm]{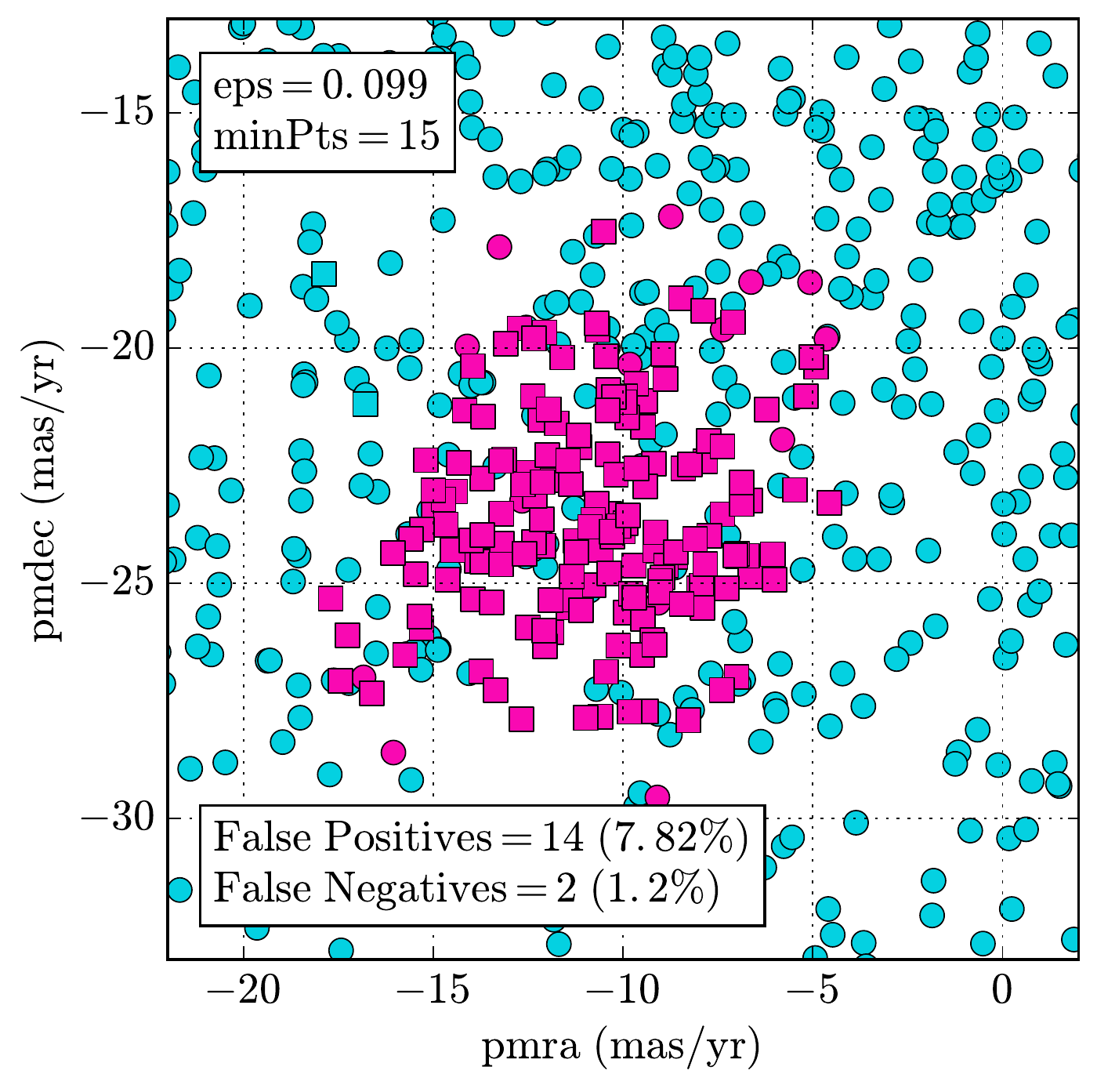} &
  \includegraphics[width=50mm]{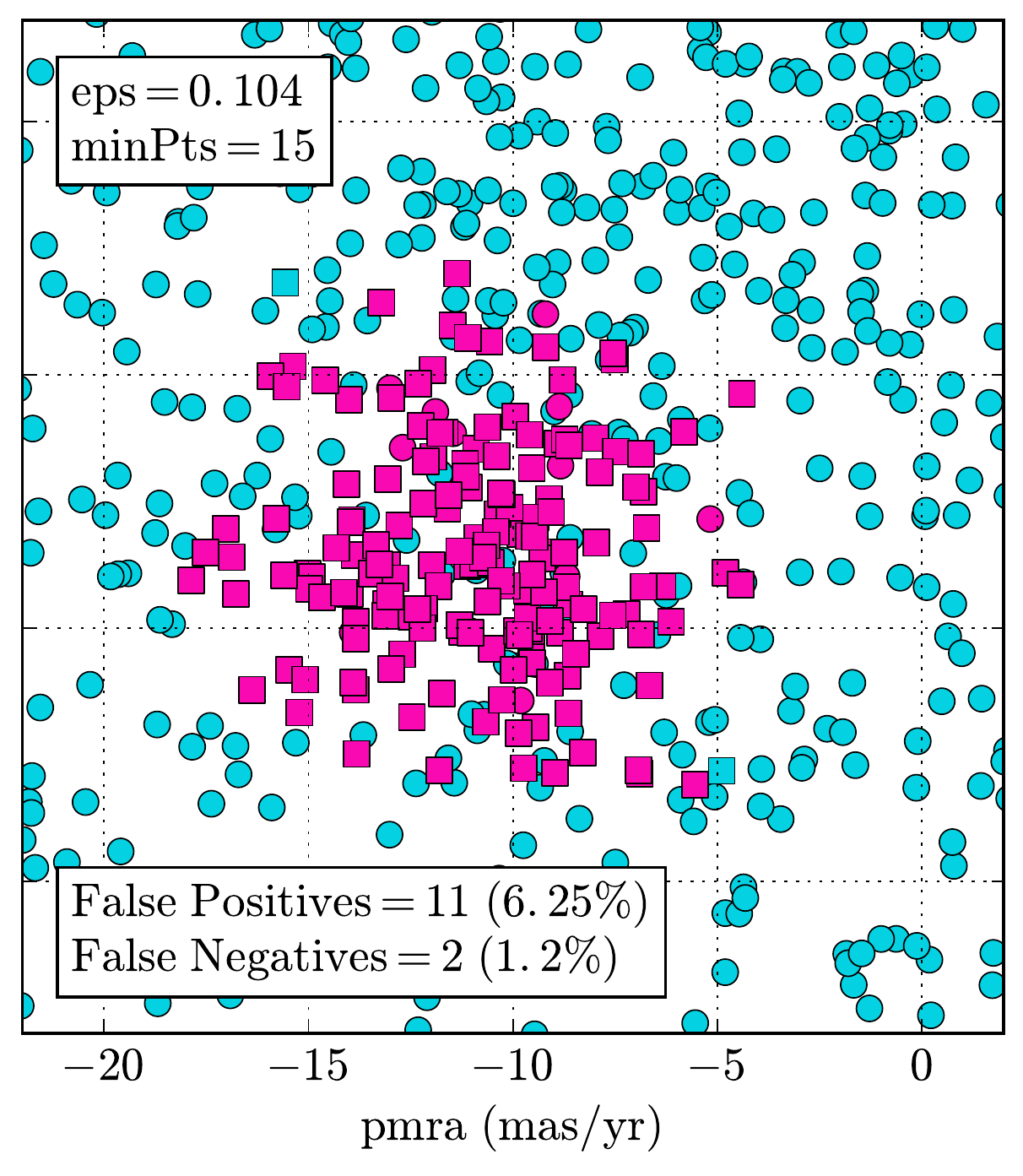} &
  \includegraphics[width=50mm]{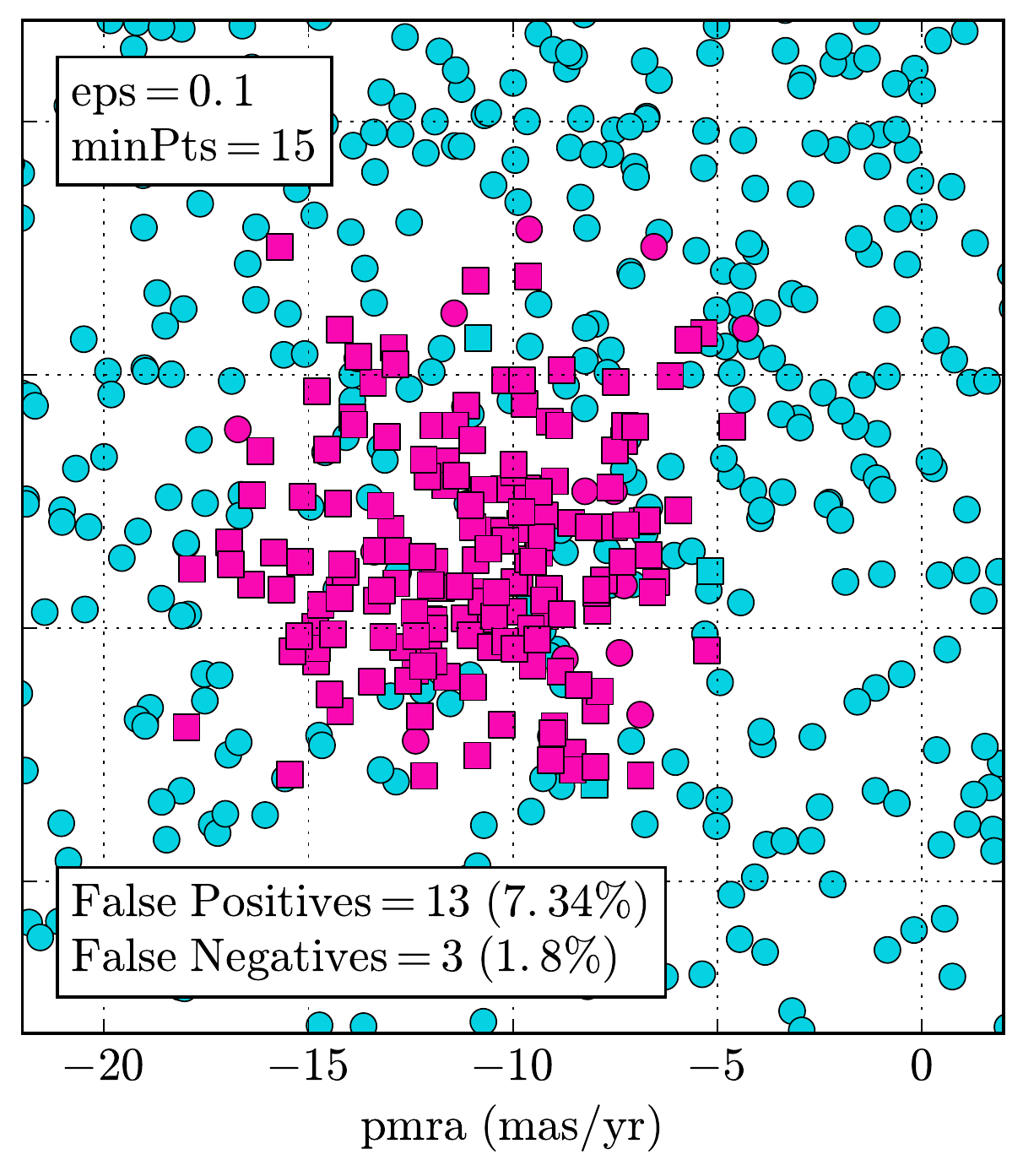}
  \end{tabular}
  \captionof{figure}{Proper motion plots showing the results of DBSCAN performed on three samples of data simulated with the binning method. True cluster members are represented by squares, and true background objects by circles. Objects selected by DBSCAN are shown in pink.}
  \label{table:bin}
\end{table*}

\begin{table*} [h!]
\centering
  \begin{tabular}{ccc}
  \includegraphics[width=57.75mm]{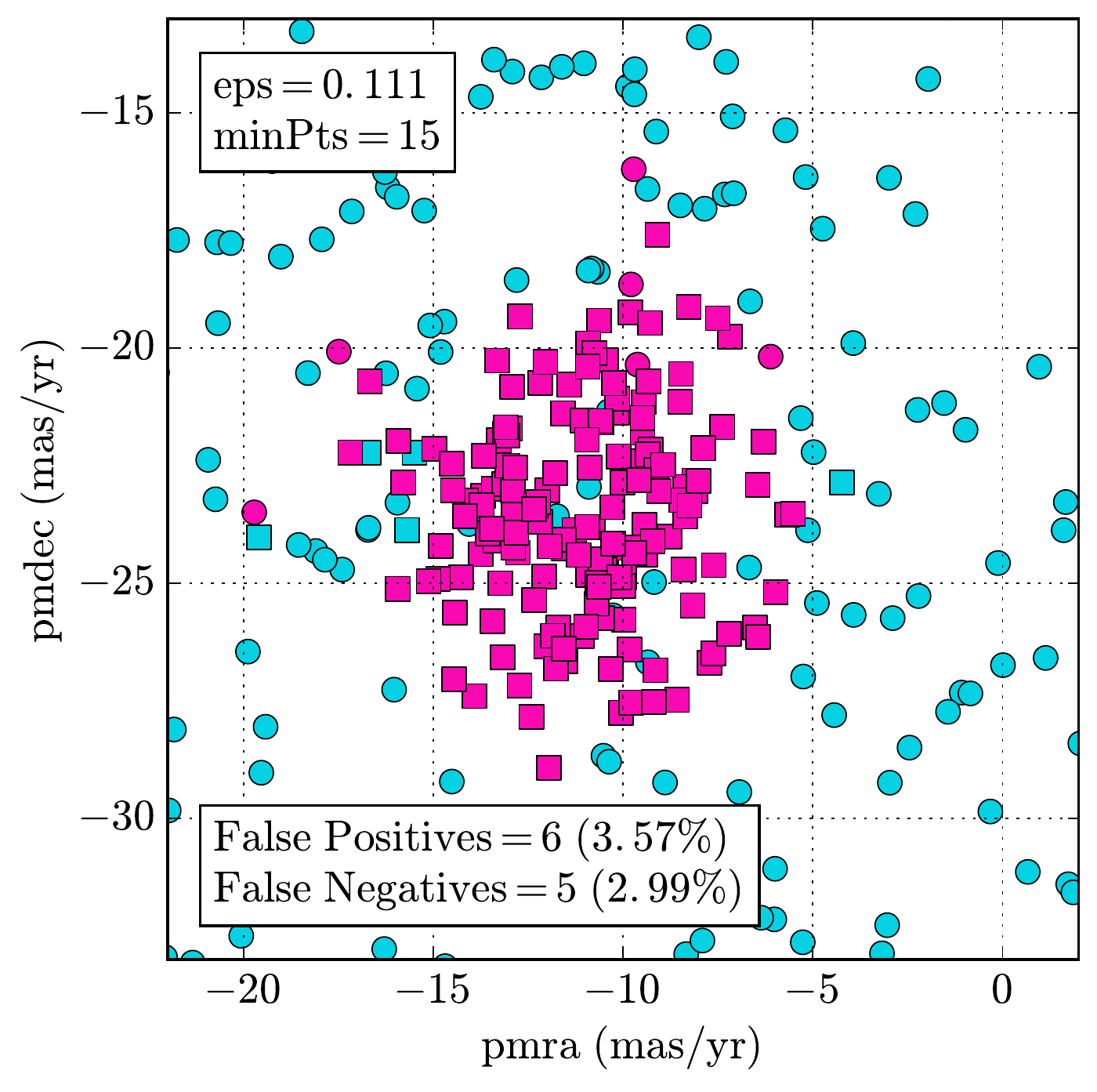} &
  \includegraphics[width=50mm]{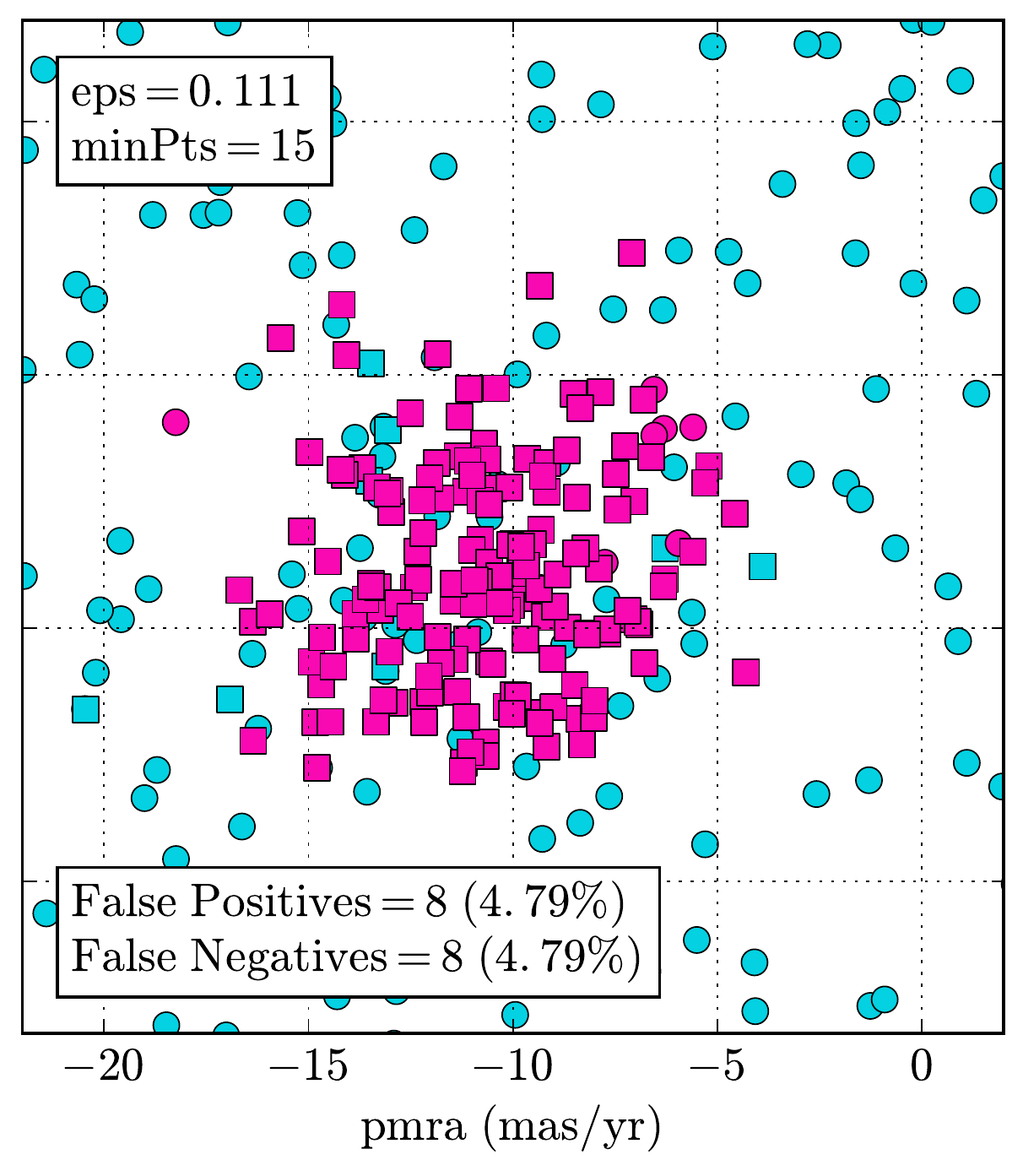} &
  \includegraphics[width=50mm]{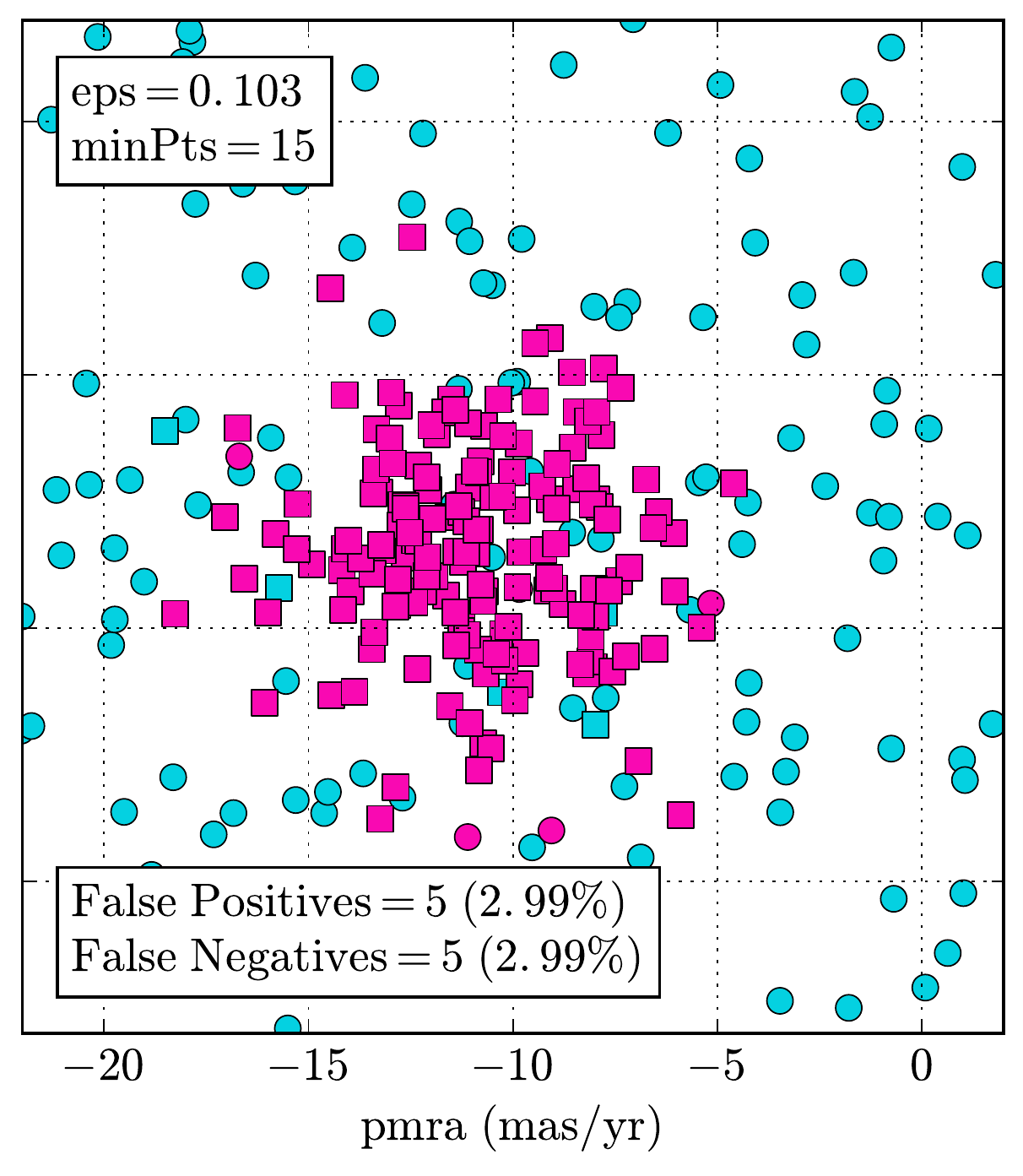}
  \end{tabular}
  \captionof{figure}{Proper motion plots showing the results of DBSCAN performed on three samples of data simulated with the Gaussian method. True cluster members are represented by squares, and true background objects by circles. Objects selected by DBSCAN are shown in pink.}
  \label{table:gauss}
\end{table*}

\addtolength{\tabcolsep}{4.5pt} 

In order to verify the suitability of the DBSCAN for the purpose of this analysis, we applied our method to two sets of simulations. One set was produced by binning the data, the other by modelling the cluster and background as Gaussian. We simulated the cluster selected by DBSCAN and the remaining background stars independently. In the binning method, for each group, the data was binned over distance and proper motion in right ascension and declination using the bin width obtained by the Rice rule \citep{jones2001} applied to the cluster. Simulated data for each bin was then generated using a random uniform distribution in each dimension. 

For each sample, a sorted k-distance graph was created, and the \textit{eps} was identified. DBSCAN was then performed on the simulated data using the identified \textit{eps} and a \textit{minPts} of 15. The cluster selected by DBSCAN from the simulated data was then analysed to determine the number of false positives (objects not in the simulated cluster, selected by DBSCAN) and false negatives (objects in the simulated cluster, not selected by DBSCAN). The percentage of false positives was calculated from the number of false positives over the size of the cluster selected by DBSCAN, and the percentage of false negatives was calculated from the number of false negatives over the size of the simulated cluster. This process was repeated 20 times for each simulation method. For the binning method, we obtained a mean false positive percentage of 10.2\% and false negative percentage of 5.1\%. For the Gaussian method, we obtained a mean false positive percentage of 3.8\% and false negative percentage of 6.0\%. Figure \ref{table:bin} and \ref{table:gauss} illustrate a sample of the results of DBSCAN performed on the data simulated using the binning method and Gaussian method respectively. 

As discussed in Section 3, and illustrated by Figure \ref{fig:pmpm_overlay}, DBSCAN can successfully recover the core of the proper motion distribution of a cluster, but struggles at the low-density extremes. Therefore for the Gaussian simulation, DBSCAN will likely recover the core of the simulated Gaussian distribution effectively, resulting in a lower false positive rate. With the binning method, since we simulate the distribution of the original member candidates, we are attempting re-select the core of the distribution directly without a surrounding distribution. Compared to the Gaussian method, the binning method will likely have a higher false positive rate. These two methods provide a possible range for the number of likely contaminants in the member candidates, with the the Gaussian simulation as a lower bound and the binning simulation as an upper bound. This gives a range of likely contaminants of $\sim$6-17 ($\sim$3.8-10.2\%).

\section{DBSCAN Parameter Sensitivity}

In order to demonstrate the behaviour of DBSCAN under the variation of the parameters, we reran DBSCAN on the data for different combinations of \textit{eps} and \textit{minPts}. Figure \ref{table:parameter} shows the results, in proper motion space, of 12 different parameter combinations. The top six plots vary the \textit{eps} between 0.5 and 0.05 for a \textit{minPts} of 15, which is the value of \textit{minPts} used in our main analysis. We see that for an \textit{eps} significantly larger than 0.103, the value determined from the k-distance graph, DBSCAN selects almost the entirety of the data as a single cluster. The cluster selected collapses inwards as the \textit{eps} is lowered towards 0.103. For an \textit{eps} of 0.05, DBSCAN selects a small portion of the core of the cluster associated with Upper Scorpius.

The bottom six plots vary \textit{minPts} to 5, 30, and 50, showing the results of DBSCAN for an \textit{eps} of 0.103, as used for a \textit{minPts} of 15, and for the value determined for each \textit{minPts} from their respective k-distance graph. We see that for a \textit{minPts} of 5, an extremely large number of clusters are selected, 33 for an \textit{eps} of 0.103 and 12 for an \textit{eps} of 0.064. This is because \textit{minPts} is too small, so DBSCAN will select small scale over-densities in the data that most likely do not correspond to any meaningful structure. For a \textit{minPts} of 30 and 50, with an \textit{eps} of 0.103, which in this case is too small a value, DBSCAN will select a subset of the cluster associated with Upper Scorpius. With the \textit{eps} selected by the k-distance graph for each case, the clusters are much closer to that obtained in our main analysis.

\addtolength{\tabcolsep}{-4.5pt}

\begin{table*}
\centering
  \begin{tabular}{ccc}
  \includegraphics[width=57.75mm]{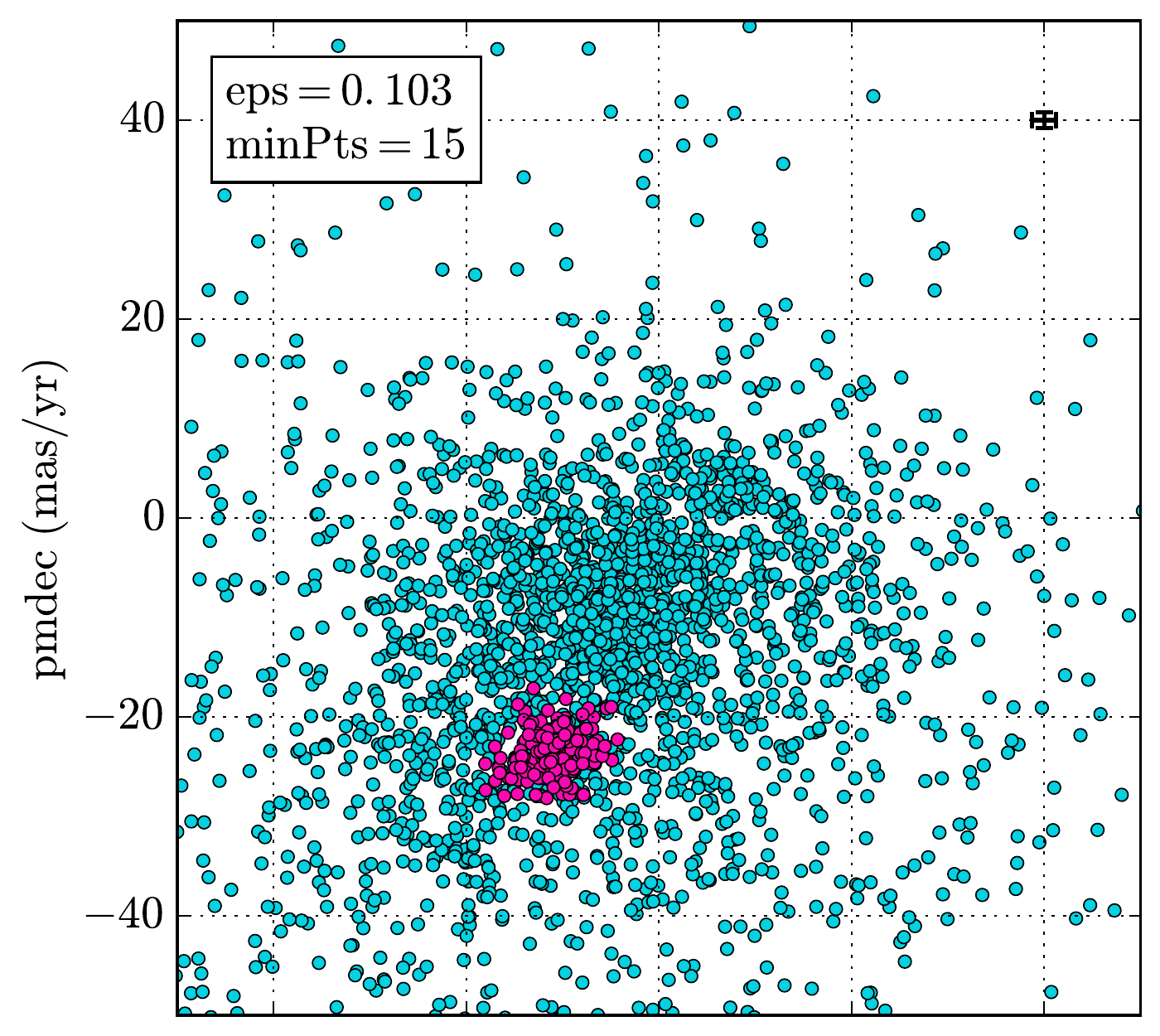} &
  \includegraphics[width=50mm]{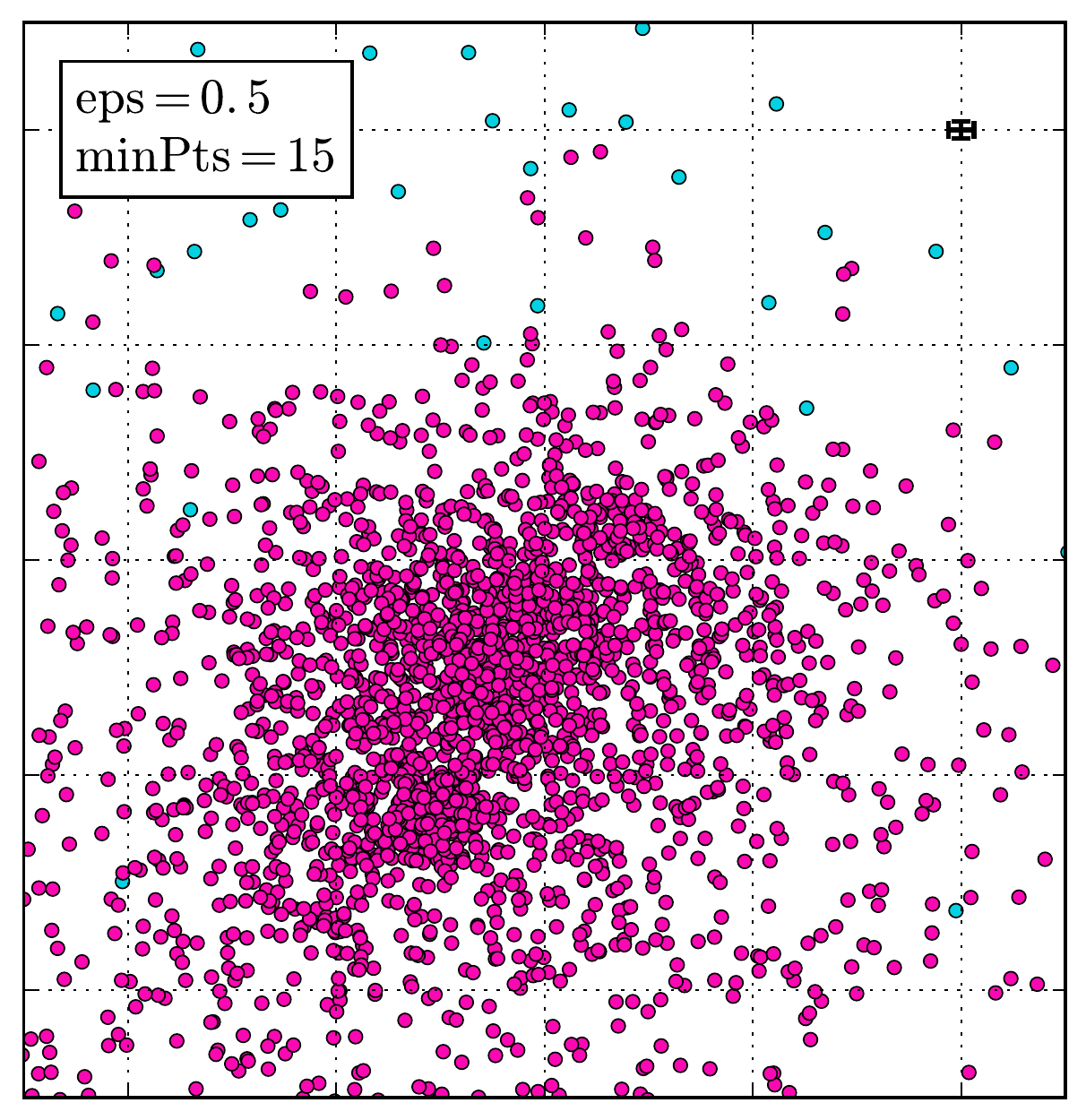} &
  \includegraphics[width=50mm]{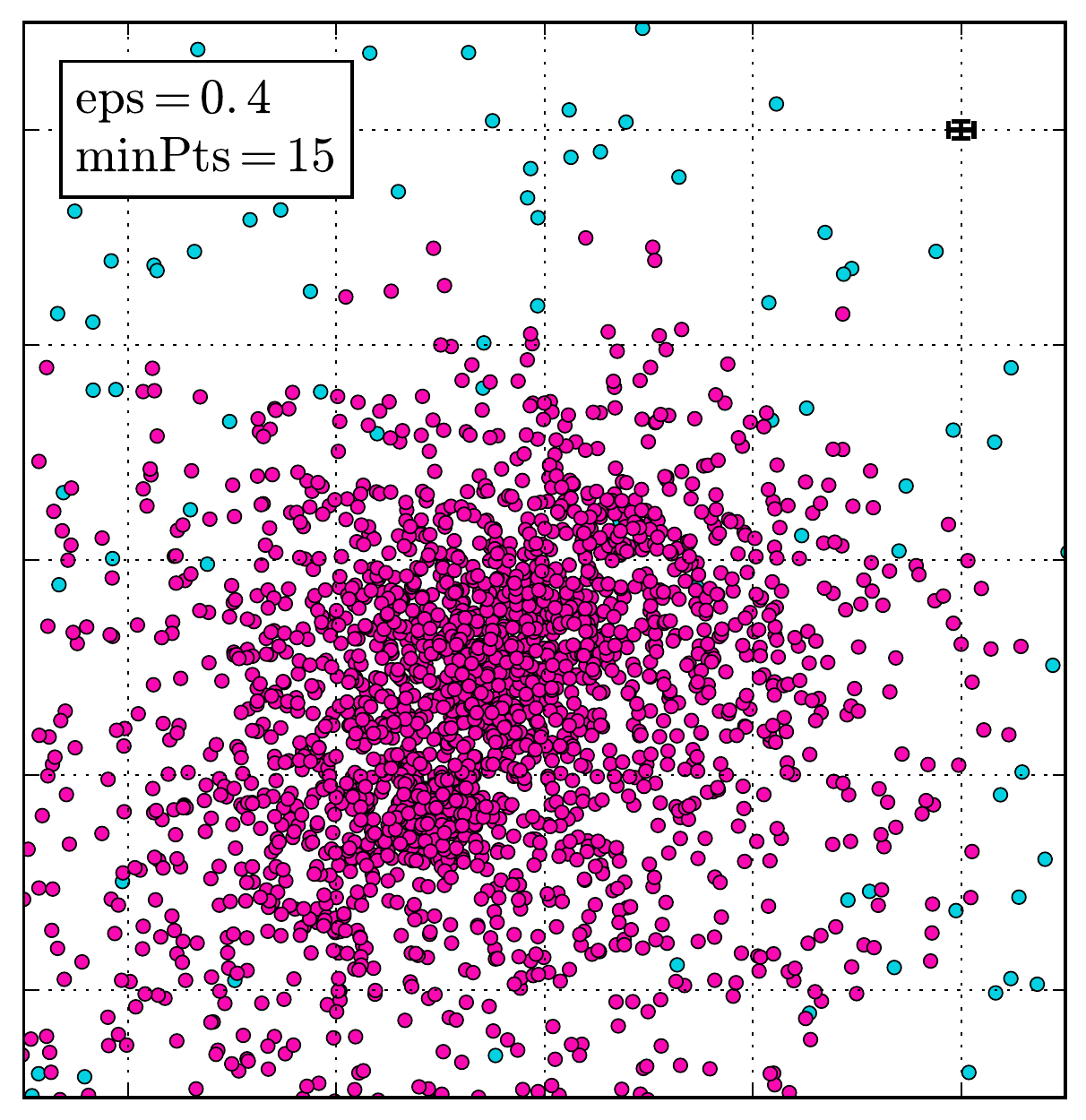}\\
  \includegraphics[width=57.75mm]{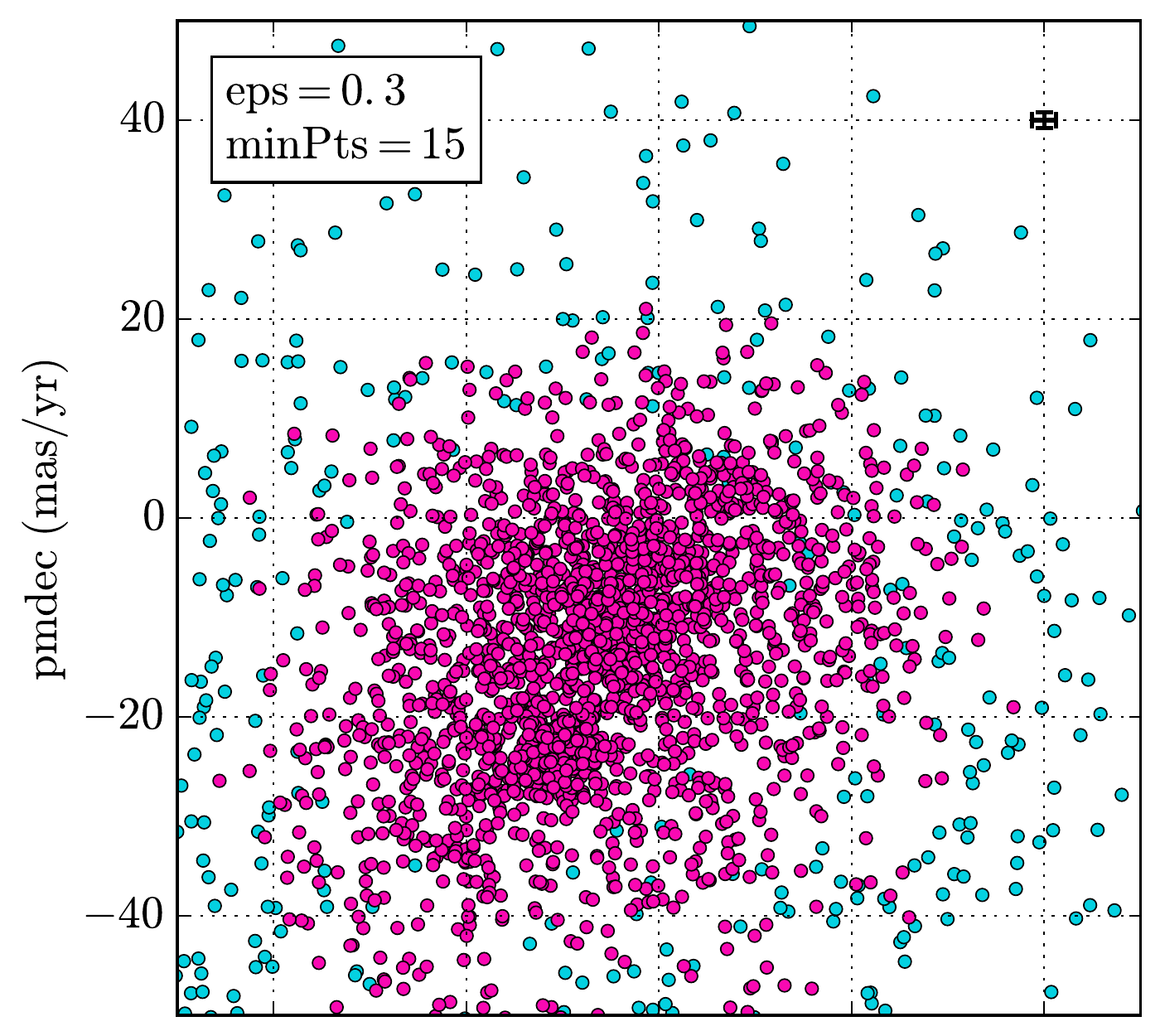} &
  \includegraphics[width=50mm]{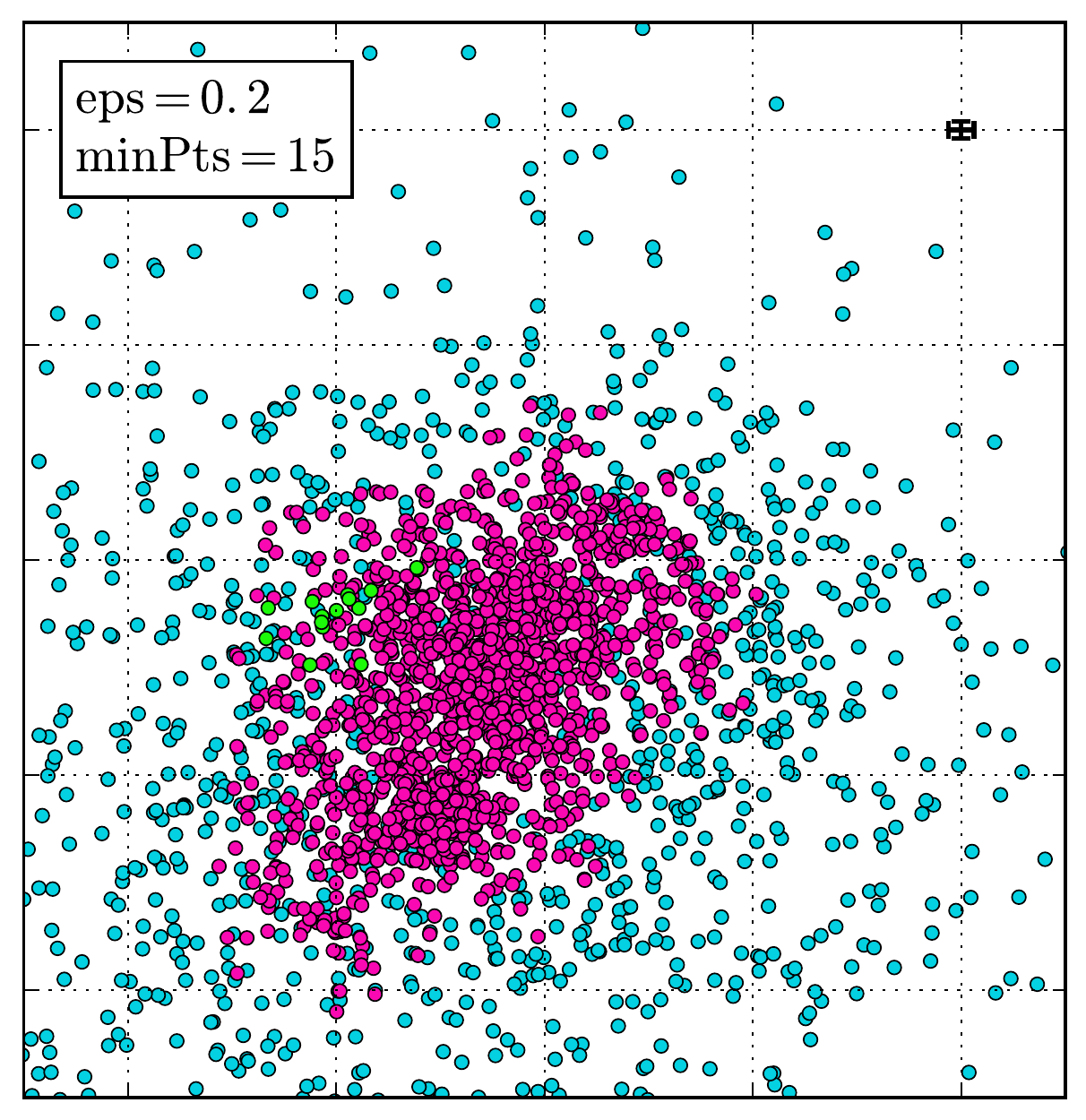} &
  \includegraphics[width=50mm]{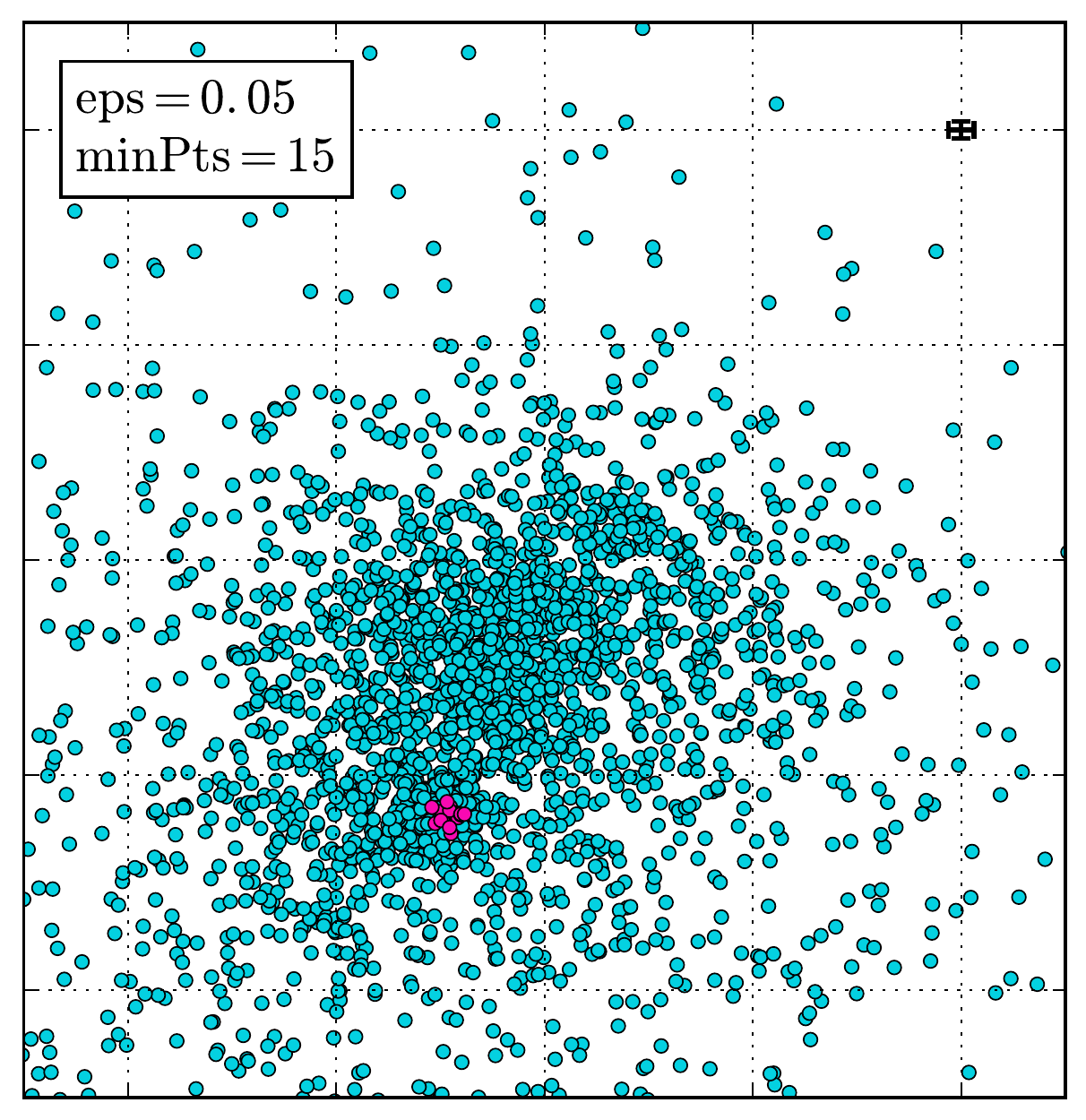}\\
  \includegraphics[width=57.75mm]{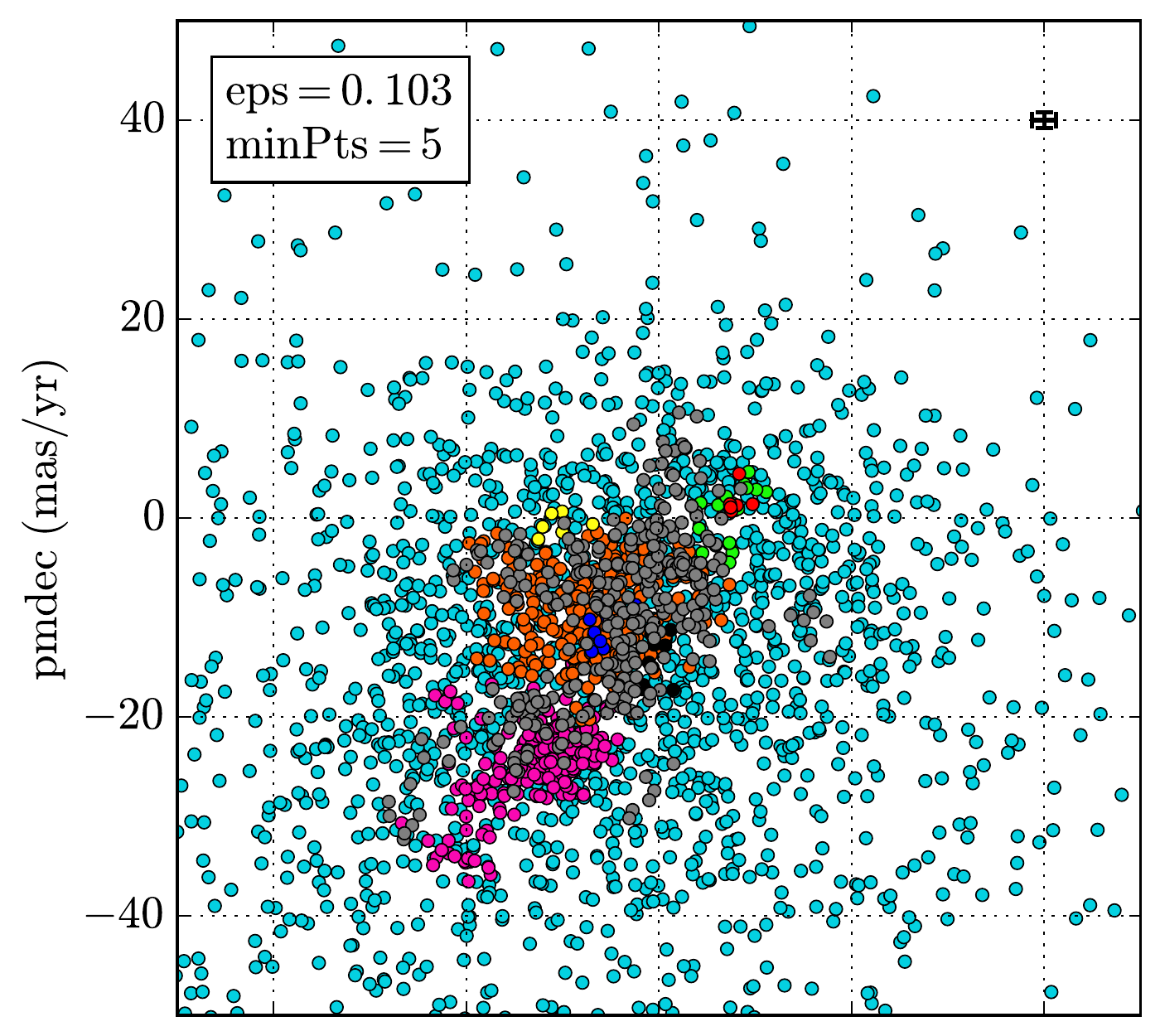} &
  \includegraphics[width=50mm]{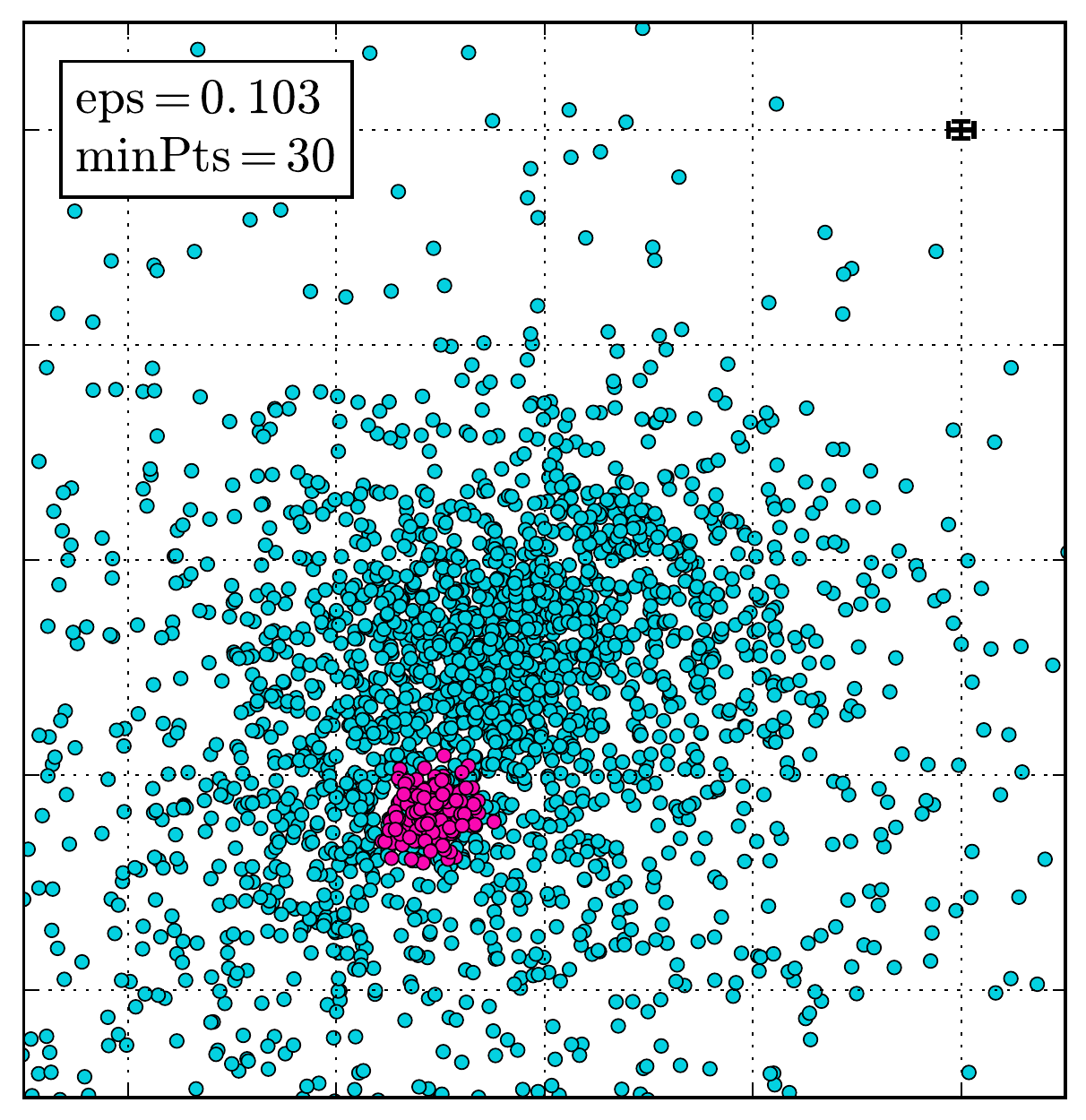} &
  \includegraphics[width=50mm]{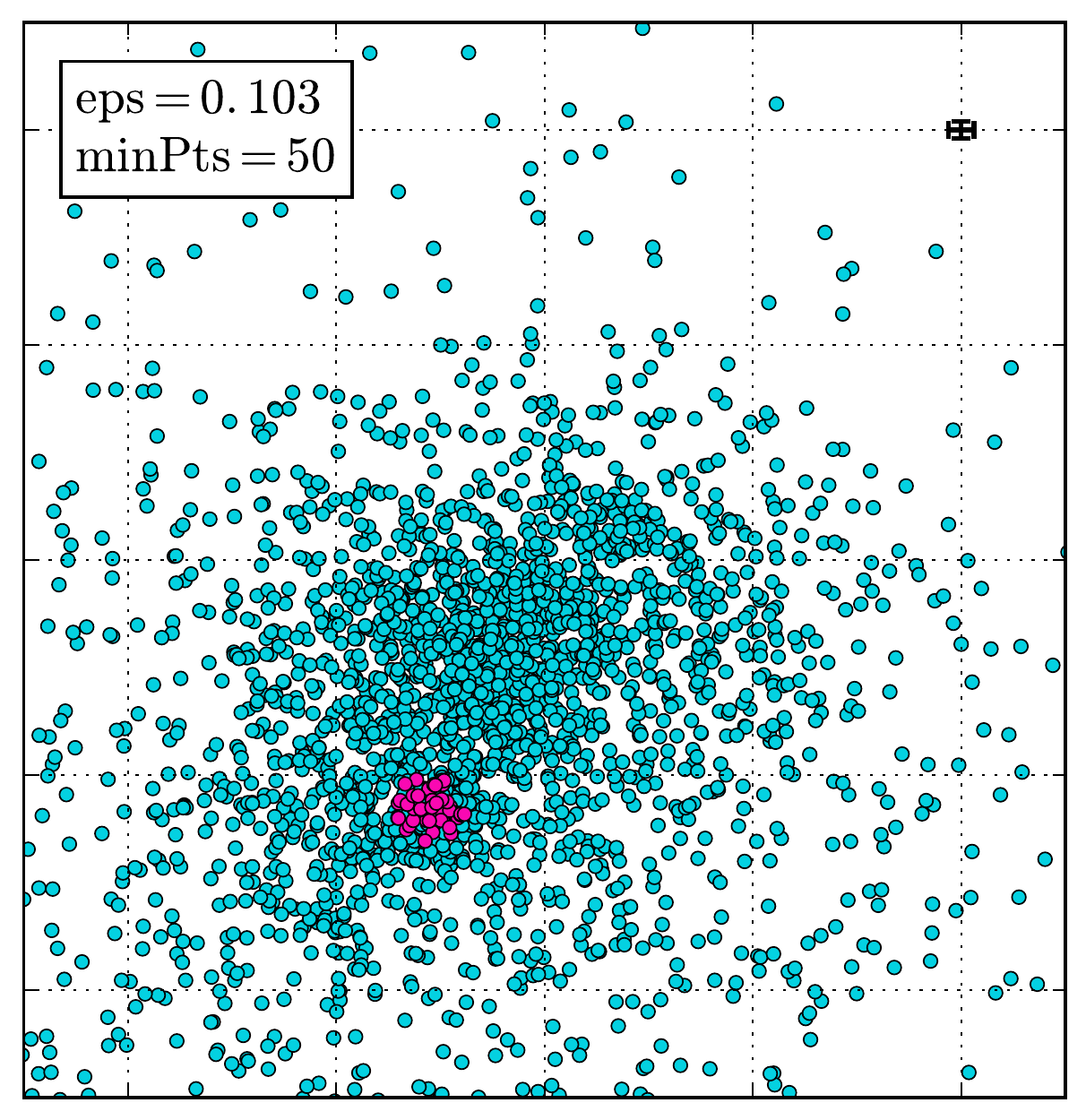}\\
  \includegraphics[width=57.75mm]{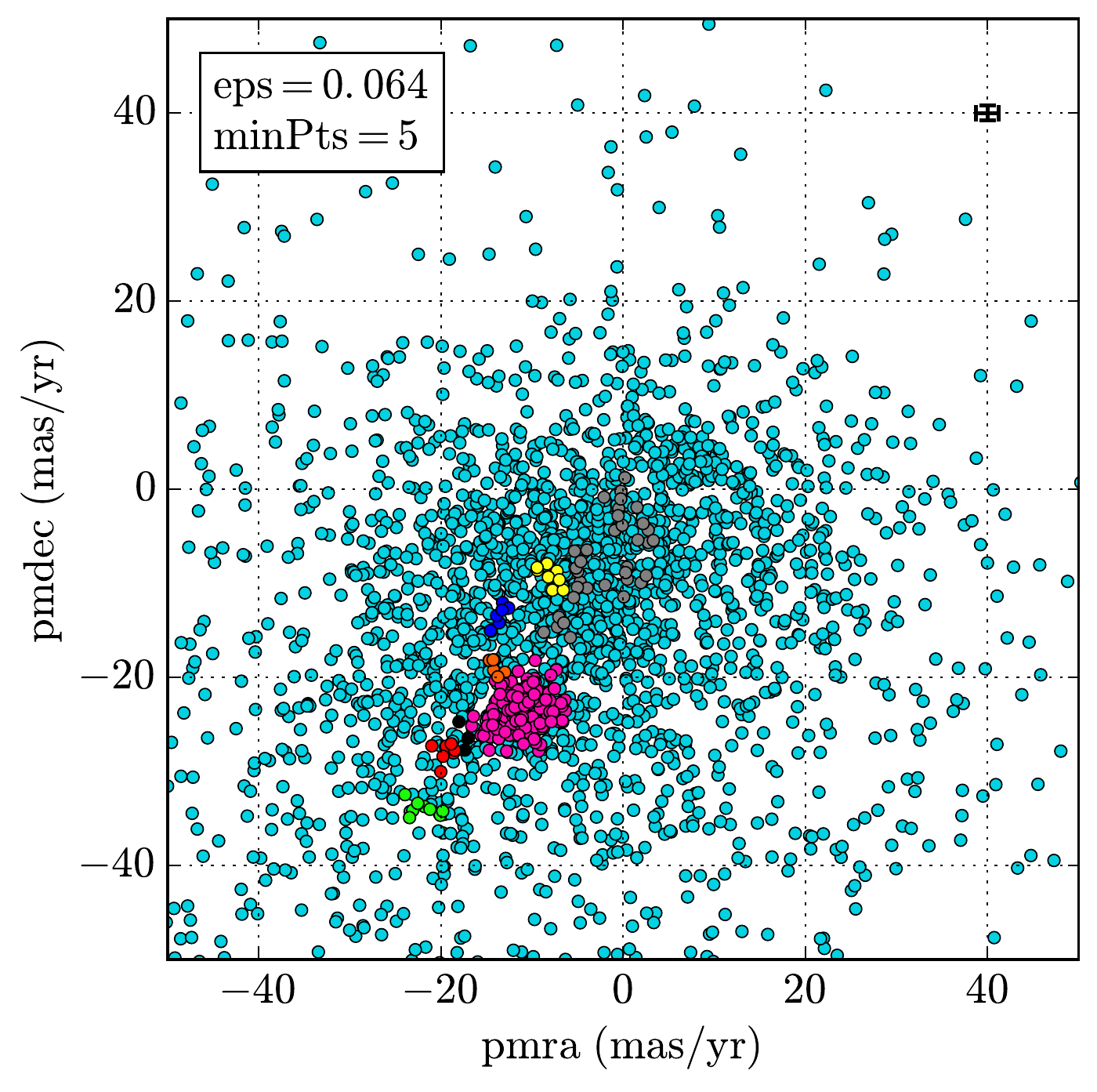} &
  \includegraphics[width=50mm]{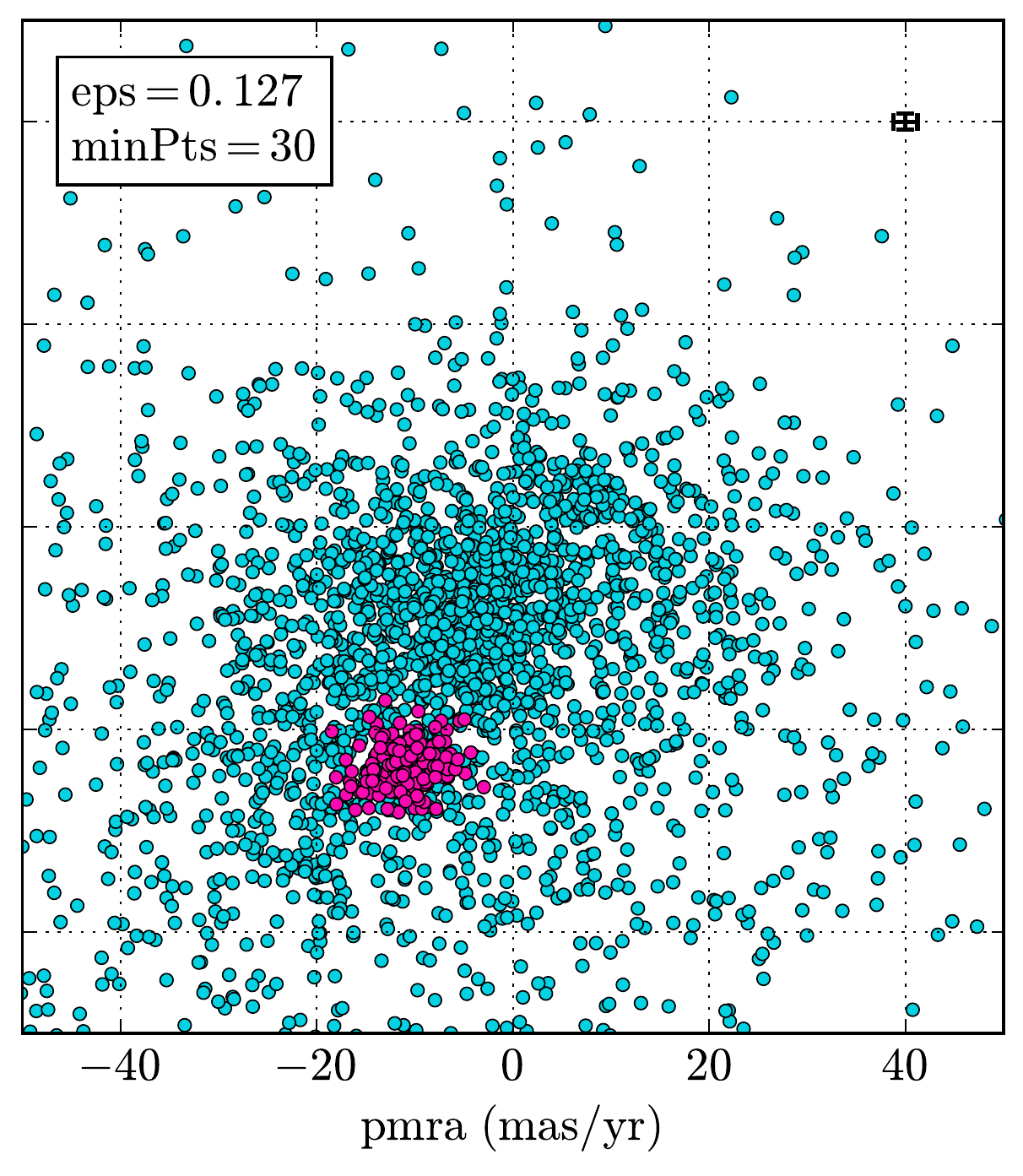} &
  \includegraphics[width=50mm]{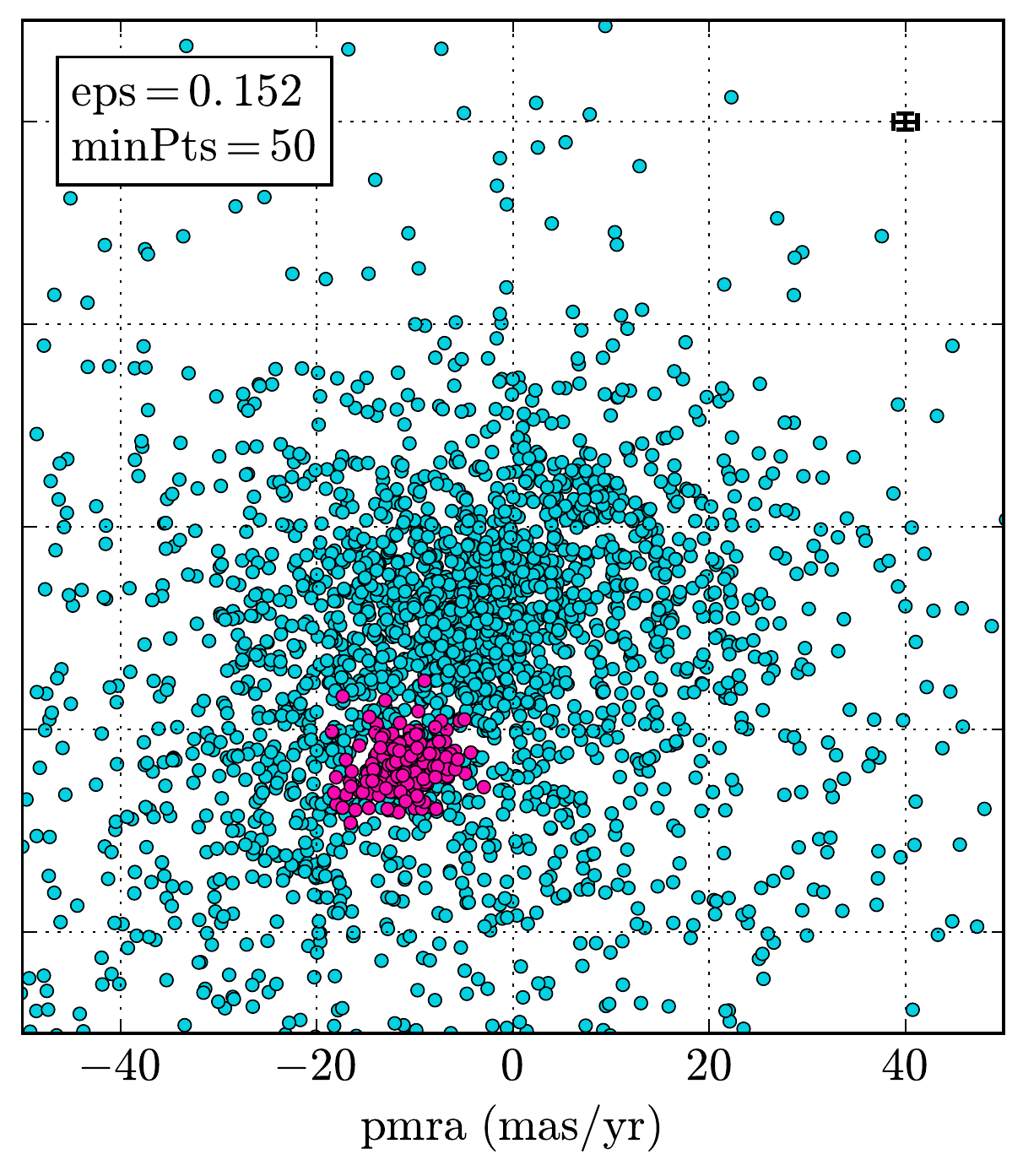}
  \end{tabular}
  \captionof{figure}{Proper motion plots showing the results of DBSCAN performed, with various combinations of parameters, on the TGAS sample. Background objects are shown in blue, with each cluster being shown in a different color. If more than 7 clusters are selected, the excess clusters are all shown in grey.}
  \label{table:parameter}
\end{table*}

\addtolength{\tabcolsep}{4.5pt}

\end{appendix}
\end{document}